\documentclass[aps,showpacs,superscriptaddress,tightenlines,a4paper]{revtex4}

\usepackage{amssymb}
\usepackage{amsmath}
\usepackage{bbold}
\usepackage{mathrsfs}
\usepackage{paralist}
\usepackage{graphicx}
\usepackage{xcolor}

\begin{document}

%
%
    \title{Polariton excitation rates from time dependent dielectrics}
%
%
	\author{S. Bugler--Lamb}
	\affiliation{Department of Physics and Astronomy, University of Exeter,
Stocker Road, Exeter, EX4 4QL}
	\email{slb235@exeter.ac.uk}
	
	\author{S. A. R. Horsley}
	\affiliation{Department of Physics and Astronomy, University of Exeter,
Stocker Road, Exeter, EX4 4QL}
	\email{s.horsley@exeter.ac.uk}
	
%
%
    \begin{abstract}
	In recent years, a rigorous quantum mechanical model for the interaction between light and macroscopic dispersive, lossy dielectrics has emerged---macroscopic QED---allowing the application of the usual methods of quantum field theory. Here, we apply time dependent perturbation theory to a general class of problems involving time dependent lossy, dispersive dielectrics. The model is used to derive polariton excitation rates in three illustrative cases, including that of a travelling Gaussian perturbation to the susceptibility of an otherwise infinite homogeneous dielectric, motivated by recent experiments on analogue Hawking radiation.  We find that the excitation rate is increased when the wave--vector and frequency of each polariton in the pair either satisfies (or nearly satisfies) the dispersion relation for electromagnetic waves, or is close to a material resonance.
    \end{abstract}

    \maketitle
%
%

\section{Introduction}
\par
For a large class of problems, the interaction between the electromagnetic field and macroscopic magneto--dielectrics is modelled by the use of a permittivity $\epsilon$ and permeability $\mu$.  In order to consider a broad range of frequencies of radiation, it becomes necessary to account for losses and dissipation in matter.  However it has proven difficult to construct a theory that accounts for the effects of dispersion and dissipation in systems where quantum mechanical behaviour is important; e.g. the Casimir effect~\cite{CasimirPlates,TomCasimir}, quantum friction~\cite{QuantumFriction, ShearingTheVacuum, NoQuantumFriction}, or even analogues of Hawking radiation~\cite{HawkingRadiation,AnalogHawking,WaterAnalog,FaccioAnalog,FaccioAnalog2}.  Although a great deal of progress can be made through applying the fluctuation--dissipation theorem~\cite{KramersKronig}, the absence of---for instance---a Hamiltonian operator can make it difficult to resolve conflicting predictions (e.g. \cite{NoQuantumFriction,FactOrFiction,UlfResponse}). 
\par
	During recent years, much progress has been made towards a rigorous quantisation scheme that accounts for the effects of dispersion and dissipation, see for example \cite{HopfieldModel,HuttnerBarnettModel,AmooshModel,ScheelQED,TomMacro,MovingMedia,ConstitutiveEquations} and references therein.  In this work we shall apply the theory presented in \cite{TomMacro}, which is a generalization of the Huttner--Barnett theory~\cite{HuttnerBarnettModel} and uses a bath of simple harmonic oscillators to account for the energy lost from the electromagnetic field.  The usual frequency domain constitutive equations $\mathbf{D}\left(\omega\right)=\epsilon_{0}\epsilon\left(\omega\right)\mathbf{E}\left(\omega\right)$ and $\mathbf{H}\left(\omega\right)=\mathbf{B}\left(\omega\right)/\mu_{0}\mu\left(\omega\right)$ are recovered from the equations of motion, with the relative permittivity $\epsilon(\omega)$ and permeability $\mu(\omega)$ obeying the Kramers-Kronig relations~\cite{KramersKronig}.  One of the significant consequences of this theory (which was also pointed out earlier by Huttner and Barnett~\cite{HuttnerBarnettModel}), is that the excitations of the electromagnetic field and medium are strongly coupled so that one must work in terms of \emph{polaritons} instead of photons: unlike photons these quanta have no set relationship between their frequency \(\omega\) and wave--vector \(\boldsymbol{k}\).  We note that the theory was initially developed for isotropic stationary media, but can be generalised to bianisotropic and moving media \cite{MovingMedia,ConstitutiveEquations}. 
\par	
	Here we shall explore the effects of the motion of a material and the time dependence of its properties on the excitation of the system when it is initially prepared in the ground state.  Although such dynamic Casimir type effects have been studied before~\cite{Dodonov2010}, one typically does not include the effects of dispersion and dissipation, which we include here within the full generality of the formalism of macroscopic QED.  Our analysis predicts the excitation rate of polaritons within the material as a function of \(\omega\) and \(\boldsymbol{k}\), and can be applied to any material so long as the permittivity and permeability satisfy the Kramers--Kronig relations.  The excitations we find do not obey a dispersion relation, but they each possess a time dependent electric field able to excite a detector embedded within the material.  They also produce radiation that could be measured by a photon detector placed outside of the material, although we do not calculate this process here.
\par
\begin{figure}[h]
		\includegraphics[height=4cm]{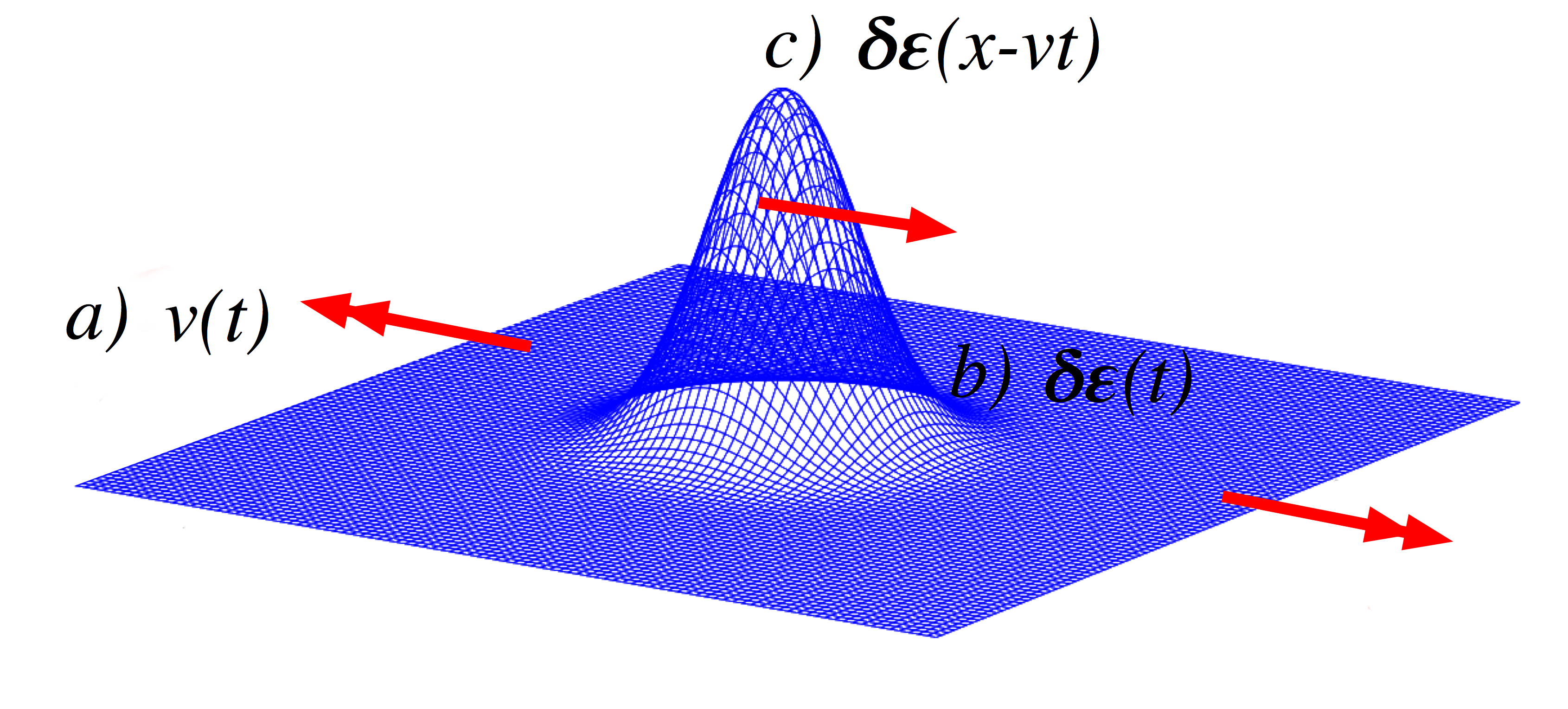}
		\caption{The three illustrative cases of macroscopic QED applied to moving and time dependent media explored in this paper: a) an infinite homogeneous medium oscillates at velocity \(\mathbf{v}(t)=v(t)\hat{\mathbf{z}}\) around a fixed position; b) the permittivity of a uniform medium oscillates around a fixed value, \(\epsilon(\omega,t)=\epsilon_{b}(\omega)+\delta\epsilon(\omega,t)\); c) a spatially dependent change to the permittivity $\delta\epsilon(\omega,x,t)=\delta\epsilon(x-vt,\omega)$, moves at a constant velocity \(\mathbf{v}=v\hat{\mathbf{z}}\) through an otherwise uniform medium of permittivity $\epsilon_{b}(\omega)$, emitting radiation as it moves.  To calculate the emission rates we apply time dependent perturbation theory to macroscopic QED~\cite{TomMacro}.\label{fig:System}}
\end{figure}

	
	In the first section of this paper we explore the classical Lagrangian and Hamiltonian necessary to describe a dispersive, lossy dielectric that has time dependent properties, or is in motion.  We briefly show that in both cases (see figure \ref{fig:System}) the effects of motion or time dependence can be subsumed within new effective material susceptibilities.  Using time--dependent perturbation theory, we then calculate the excitation rate of pairs of polaritons due to the time dependence or motion of the medium.   We find that from the perspective of macroscopic QED time dependent material properties and material motion contribute separately to the Hamiltonian, even though one would expect a moving spatial distribution of permittivity to be indistinguishable from a moving medium.  After developing the general theory we treat three illustrative cases (see figure~\ref{fig:System}):  (a) an infinite homogeneous medium performing a linear oscillatory motion; (b) an	 infinite homogeneous medium where the permittivity oscillates in time; and (c) a `bump' in the permittivity moving uniformly through an otherwise homogeneous dielectric.  The final example is inspired by the recent optical experiments investigating laboratory analogues of Hawking radiation~\cite{AnalogHawking,FaccioAnalog}.
	
\section{Lagrangian of electromagnetism in a time dependent, moving dielectric}
\par	
	In this section we construct the classical theory of the electromagnetic field interacting with a moving or time dependent material that exhibits dispersion and dissipation.  The Lagrangian of \cite{TomMacro} is extended in a similar fashion to~\cite{MovingMedia,ConstitutiveEquations}, in this case accounting for non-relativistic motion and an arbitrary time dependence of the permittivity profile.  The dynamical variables of the electromagnetic field are the vector potential $\mathbf{A}$ and the scalar potential $\phi$, which are related to the electric and magnetic field strengths by $\mathbf{E}=-\nabla\phi-\partial_{t}\mathbf{A}$ and $\mathbf{B}=\nabla\times\mathbf{A}$ respectively.  The dissipation of electromagnetic energy is governed by the coupling of the electromagnetic field to a field of simple harmonic oscillators \(\mathbf{X}_{\omega}(\mathbf{x},t)\) present throughout space, with every possible natural frequency \(\omega\).   The Lagrangian density for our system is given by
	
\begin{equation}\label{eq:LagrDens}
\begin{split}	
	\mathscr{L}=\frac{\epsilon_{0}}{2}\left(\mathbf{E}^{2}-c^{2}\mathbf{B}^{2}\right)&+\mathbf{E}\cdot\intop_{0}^{\infty}d\omega\ \alpha\left(\omega,\mathbf{x},t\right)\cdot\mathbf{X}_{\omega}\\&+\mathbf{B}\cdot\intop_{0}^{\infty}d\omega\  \boldsymbol{\alpha_{B}}\left(\omega,\mathbf{x},t\right)\cdot\mathbf{X}_{\omega}+\frac{1}{2}\intop_{0}^{\infty}d\omega\left[\left(\mathbf{\dot{X}}_{\omega}-\left(\mathbf{v}(\mathbf{x},t)\cdot\mathbf{\nabla}\right)\mathbf{X}_{\omega}\right)^{2}-\omega^{2}\mathbf{X}_{\omega}^{2}\right]
\end{split}
\end{equation}
and is valid to first order in the velocity $\mathbf{v}(\mathbf{x},t)$ of the material, which at this point we allow to be spatially dependent.  For simplicity we have set $\mu=1$ throughout, which amounts to ignoring the second oscillator bath that appears in~\cite{TomMacro,MovingMedia,ConstitutiveEquations}, although this can be re--introduced with no fundamental modifications to our results.  The coupling between the field of oscillators and the electromagnetic field is mediated via a term that is proportional to a quantity $\alpha$ which in this case is a scalar quantity that depends on space, time and frequency.  The second coupling term $\boldsymbol{\alpha_{B}}(\omega,\mathbf{x},t)$ is given in terms of \(\alpha\) by $\boldsymbol{\alpha_{B}}\left(\omega,\mathbf{x},t\right)=-\mathbf{v}\left(\mathbf{x},t\right)\times\alpha\left(\mathbf{x},\omega,t\right)\mathbb{1}_{3}$, and is due to the Lorentz transformation of the field strengths between reference frames (a phenomenon we shall often refer to as polarization mixing).  Here we treat the time dependence of the medium as a perturbation to an otherwise isotropic, time independent background: $\alpha\left(\omega,\mathbf{x},t\right)=\alpha_{b}\left(\mathbf{x},\omega\right)+\delta\alpha\left(\omega,\mathbf{x},t\right)$.  The background permittivity of the medium \(\epsilon_{b}(\mathbf{x},\omega)\) is related to this time independent coupling term by the expression given in \cite{TomMacro}: $\alpha_{b}\left(\mathbf{x},\omega\right)=\sqrt{2\omega\epsilon_{0}{\rm Im}[\epsilon_{b}\left(\mathbf{x},\omega\right)]/\pi}$. 
\par
The Hamiltonian of the system is found in terms of the canonical momenta and the Lagrangian (see for example \cite{QFT}) and is given by
\begin{equation}
H=\intop_{-\infty}^{\infty}d^{3}\mathbf{x}\left[\Pi_{\mathbf{\mathbf{A}}}\cdot\mathbf{\dot{A}} +\intop_{0}^{\infty}d\omega\ \Pi_{\mathbf{X_{\omega}}}\cdot\mathbf{\dot{X}_{\omega}}-\mathscr{L} \right].
\end{equation}
where the canonical momenta of the electromagnetic field $\Pi_{\mathbf{\mathbf{A}}}$ and of the oscillator field $\Pi_{\mathbf{X_{\omega}}}$ are given by
 \begin{equation}
 	\Pi_{\mathbf{\mathbf{A}}}=-\epsilon_{0}\mathbf{E}-\intop_{0}^{\infty}d\omega\ \boldsymbol{\alpha}\left(\omega,\mathbf{x},t\right)\cdot\mathbf{X}_{\omega}
 \end{equation}
 \begin{equation}
 	\Pi_{\mathbf{X_{\omega}}}=\dot{\mathbf{X}}_{\omega}-\left(\mathbf{v}\cdot\nabla\right)\mathbf{X}_{\omega}
 \end{equation}
The scalar potential $\phi$ is not a dynamical variable, and can be removed from the Hamiltonian as is done in e.g. \cite{TomMacro}. 
\par
Because we treat the time dependence of the dielectric as having a small effect on the total field, we divide this Hamiltonian up into two parts,
\begin{equation}\label{HSplit}
H=H_{0}+H_{I}
\end{equation}
where $H_{0}$ is the Hamiltonian of a time-independent dielectric, similar to that of \cite{TomMacro} but with $\mu=1$, given by

\begin{equation}\label{eq:H0}
H_{0}=\frac{1}{2}\int d^{3}\mathbf{x}\left[ \frac{1}{\epsilon_{0}}\left(\Pi_{\mathbf{A}}+\intop_{0}^{\infty}d\omega\ \alpha_{b}\left(\mathbf{x},\omega\right)\mathbf{X}_{\omega}\right)^{2}+\frac{1}{\mu_{0}}\left(\nabla\times\mathbf{A}\right)^{2}+\intop_{0}^{\infty}d\omega\left( \Pi_{\mathbf{X_{\omega}}}^{2}+\omega^{2}\mathbf{X}_{\omega}^{2}\right) \right] 
 \end{equation}
 
and $H_{I}$ is the interaction Hamiltonian due to the motion and time dependence of the dielectric, given by,

\begin{equation}\label{eq:Pert}
H_{I}=\int d^{3}\mathbf{x}\intop_{0}^{\infty}d\omega\left[ \Pi_{\mathbf{X_{\omega}}}\cdot\left(\mathbf{v}\cdot\mathbf{\nabla}\right)\mathbf{X}_{\omega}
+\mathbf{B}\cdot\left( \boldsymbol{\alpha_{B}}\left(\omega,\mathbf{x},t\right)\cdot\mathbf{X}_{\omega}\right)
+\mathbf{E}\cdot \mathbf{X}_{\omega}\delta \alpha\left(\omega,\mathbf{x},t\right) \right].
\end{equation} 

The interaction Hamiltonian $H_{I}$ contains three terms: the first accounts for the fact that the bath of oscillators is in motion; the second is due to the polarization mixing due to the motion; and the third term is due to the time dependences of the permittivity, which could---for example---be due to time dependent boundaries or simply the time-varying permittivity of a stationary object.
\par
	Before embarking on the quantum mechanical calculation, we derive the classical equations of motion and show how the time dependence of the system can be seen as a modification of the relative permittivity.  As an example, consider the case of a spatially homogeneous medium moving with an arbitrary time dependent velocity, where \(\delta\alpha=0\).   Through the use of the Euler-Lagrange equations (see e.g. \cite{QFT}), we obtain the equations of motion for the oscillator bath 

\begin{equation}\label{eq:OscillatorEquationsOfMotion}
\left[\left(\frac{\partial}{\partial t}+\left(\mathbf{v}\cdot{\nabla}\right)\right)^{2}+\omega^{2}\right]\mathbf{X}_{\omega}\left(\mathbf{x},t\right)=\alpha_{b}\left(\omega\right)\mathbf{E}\left(\mathbf{x},t\right)
\end{equation}

and for the field
\begin{equation}\label{eq:EWaveEQ}
\nabla\times\nabla\times\mathbf{E}+\frac{1}{c^{2}}\frac{\partial^{2}\mathbf{E}}{\partial t^{2}}=-\mu_{0}\frac{\partial^{2}\mathbf{P}}{\partial t^{2}}
\end{equation} 

with the polarisation $\mathbf{P}$ given by
\begin{equation}\label{Polarisation}
\mathbf{P}\left(\mathbf{x},t\right)=\intop_{0}^{\infty}d\omega\ \alpha_{b}\left(\omega\right)\mathbf{X}_{\omega}\left(\mathbf{x},t\right).
\end{equation}

For simplicity we have ignored the polarisation mixing term $\mathbf{B}\cdot\intop_{0}^{\infty}d\omega\  \boldsymbol{\alpha_{B}}\left(\omega,\mathbf{x},t\right)\cdot\mathbf{X}_{\omega}$, concentrating on only the electric response of the material.  In order to see how the time dependence of the medium affects the material parameters, we solve equations (\ref{eq:OscillatorEquationsOfMotion}) and (\ref{eq:EWaveEQ}).  The solutions to (\ref{eq:OscillatorEquationsOfMotion}) are given by
\begin{equation}\label{eq:OscillatorSolutions}
\mathbf{X}_{\omega}\left(\mathbf{x},t\right)=\alpha_{b}\left(\omega\right)\intop d^{3}x'\intop_{-\infty}^{\infty}dt'\ G_{X_{\omega}}\left(\mathbf{x},\mathbf{x'},t,t'\right)\mathbf{E}\left(\mathbf{x},t\right)+\mathbf{X}^{H}_{\omega}\left(\mathbf{x},t\right)
\end{equation} 
where the $\mathbf{X}^{H}_{\omega}\left(\mathbf{x},t\right)$ are the homogeneous solutions to (\ref{eq:OscillatorEquationsOfMotion}) and are of the general form
\begin{equation}
	\mathbf{X}^{H}_{\omega}\left(\mathbf{x},t\right)=\mathbf{h}_{\omega}\left(\mathbf{x}+\int_{t_{0}}^{t}\mathbf{v}\left(t'\right)dt'\right)e^{-i\omega t}+c.c
\end{equation}
where $\mathbf{h}_{\omega}$ are arbitrary functions of position and the lower limit in the integral, $t_{0}$ is  the initial time of the evolution of the system where the amplitudes are simply $\mathbf{h}_{\omega}\left(\mathbf{x}\right)$.  In the quantum theory, these amplitudes will become the creation and annihilation operators of the quantum fields in the same way as the theory developed in \cite{TomMacro}. The oscillator Green's function $G_{X_{\omega}}\left(\mathbf{x},\mathbf{x'},t,t'\right)$ in (\ref{eq:OscillatorSolutions}) satisfies
\begin{equation}
	\left[\left(\frac{\partial}{\partial t}+\left(\mathbf{v}\cdot\mathbf{\nabla}\right)\right)^{2}+\omega^{2}\right]G_{X_{\omega}}\left(\mathbf{x},\mathbf{x'},t,t'\right)=\delta^{(3)}\left(\mathbf{x}-\mathbf{x'}\right)\delta\left(t-t'\right)
\end{equation}
the retarded solution (zero for \(t<t'\)) to which is given by
\begin{equation}\label{eq:OscillatorGreen}
	G_{X_{\omega}}\left(\mathbf{x},\mathbf{x'},t,t'\right)=\frac{1}{\omega}\theta\left(t-t'\right)\sin\left[\omega\left(t-t'\right)\right]\delta^{(3)}\left(\mathbf{x}+\intop_{t'}^{t}\mathbf{v}\left(t_{1}\right)dt_{1}\right).
\end{equation}
Substituting (\ref{eq:OscillatorGreen}) into (\ref{eq:OscillatorSolutions}) gives an expression for the amplitudes of the oscillator field in terms of the field $\mathbf{E}$ and the arbitrary amplitudes $\mathbf{h}_{\omega}$,
\begin{equation}\label{Oscill2}
	\mathbf{X}_{\omega}\left(\mathbf{x},t\right)=\frac{\alpha_{b}\left(\omega\right)}{\omega}\intop_{-\infty}^{t}dt'\sin\left[\omega\left(t-t'\right)\right]\mathbf{E}\left(\mathbf{x}+\intop_{t'}^{t}\mathbf{v}\left(t_{1}\right)dt_{1},t'\right)+\mathbf{X}^{H}_{\omega}\left(\mathbf{x},t\right).
\end{equation}
The meaning of this integral expression is simply that the oscillator amplitudes at $\mathbf{x}$ no longer depend solely on the strength of the field at $\mathbf{x}$ (as it does in~\cite{TomMacro}) but also on the previous positions of the moving oscillator $\mathbf{x}+\intop_{t'}^{t}\mathbf{v}\left(t_{1}\right)dt_{1}$ for times $t'$ \emph{before} $t$. The motion of the material thus leads to a non--local response, due to the fact that energy dissipated from the electromagnetic field is carried away from the point where it was absorbed.  The wave equation for the electric field is found through substituting expression (\ref{Oscill2}) into (\ref{Polarisation}), and combining the resultant expression with (\ref{eq:EWaveEQ})
\begin{equation}\label{eq:waveEquation}
	\nabla\times\nabla\times\mathbf{E}+\frac{1}{c^{2}}\frac{\partial^{2}\mathbf{E}}{\partial t^{2}}+\frac{1}{c^{2}}\frac{\partial^{2}}{\partial t^{2}}\int d^{3}x'\intop_{-\infty}^{\infty}dt'\ \chi\left(\mathbf{x},\mathbf{x'},t,t'\right)\mathbf{E}\left(\mathbf{x'},t'\right)=-\mu_{0}\mathbf{j}\left(\mathbf{x},t\right)
\end{equation} 
where the non-local effective susceptibility of the system $\chi$ is given by
\begin{equation}
	\chi\left(\mathbf{x},\mathbf{x'},t,t'\right)=\frac{1}{\epsilon_{0}}\intop_{0}^{\infty}d\omega\frac{\alpha^{2}_{b}\left(\omega\right)}{\omega}\ \delta^{(3)}\left(\mathbf{x}+\intop_{t'}^{t}\mathbf{v}\left(t_{1}\right)dt_{1}-\mathbf{x'}\right)\theta\left(t-t'\right)\sin\left[\omega\left(t-t'\right)\right].\label{movsuc}
\end{equation}
The theta function above ensures that the effective susceptibility is non-zero only for times past, enforcing causality and consequently also the Kramers--Kronig relations. In the limit $v\to0$, the wave equation (\ref{eq:waveEquation}) reduces to the usual stationary wave equation in dispersive media (see e.g. \cite{KramersKronig}), and (\ref{movsuc}) reduces to the local expression \(\chi(\mathbf{x},\mathbf{x'},t,t')=\delta^{(3)}(\mathbf{x}-\mathbf{x}')\chi(t-t')\), where \(\chi(t-t')\) is the Fourier transform of \(\epsilon_{b}(\omega)-1\).  The source of the electromagnetic field $\mathbf{j}(\mathbf{x},t)$ that appears in the wave equation (\ref{eq:EWaveEQ}) depends on the arbitrary functions $\mathbf{h}_{\omega}$ and is given by the non-local expression~\footnote{Note that here the source term \(\mathbf{j}\) does not have units of electric current.}
\begin{equation}\label{Current}
	\mathbf{j}\left(\mathbf{x},t\right)=\intop_{0}^{\infty}d\omega\ \alpha_{b}\left(\omega\right)\frac{\partial^{2}}{\partial t^{2}}\left[ \mathbf{h}_{\omega}\left(\mathbf{x}+\int_{t_{0}}^{t}\mathbf{v}\left(t'\right)dt'\right)e^{-i\omega t}+c.c\right].
\end{equation}
This current has the same interpretation as in \cite{TomMacro}: the un-driven part of the motion of the bath of oscillators is the source of the electromagnetic field in an absorbing medium. In \cite{TomMacro}, the $\mathbf{h}_{\omega}$ are related to the amplitude of the current \(\mathbf{j}\) in a local manner.  Here, the relative motion of the medium $\mathbf{v}$ means that the current at $\mathbf{x}$ now depends on the $\mathbf{h}_{\omega}$ at $\mathbf{x}+\int_{t_{0}}^{t}\mathbf{v}\left(t'\right)dt'$.  The nature of the non-locality in both the susceptibility (\ref{movsuc}) and the current (\ref{Current}) depends upon the motion.  In analogy to \cite{TomMacro}, the solution to (\ref{eq:waveEquation}) written in terms of the electromagnetic Green's function
\begin{equation}
	\mathbf{E}\left(\mathbf{x_{1}},t_{1}\right)=-\mu_{0}\intop d^{3}\mathbf{x}_{2}\intop_{-\infty}^{\infty}dt_{2}\ \mathbf{G}\left(\mathbf{x_{1}},\mathbf{x_{2}},t_{1},t_{2}\right)\cdot\mathbf{j}\left(\mathbf{x_{2}},t_{2}\right)
\end{equation}
where $\mathbf{G}$ is a bi-tensor satisfying
\begin{equation}\label{eq:GreensFunctionWaveEquation}
\begin{split}
	\nabla\times\nabla\times\mathbf{G}\left(\mathbf{x_{1}},\mathbf{x_{2}},t_{1},t_{2}\right)+\frac{1}{c^{2}}\frac{\partial^{2}}{\partial t_{1}^{2}}\mathbf{G}\left(\mathbf{x_{1}},\mathbf{x_{2}},t_{1},t_{2}\right)
+\frac{1}{c^{2}}\frac{\partial^{2}}{\partial t_{1}^{2}}\intop_{-\infty}^{t_{1}}dt_{3}\ \chi\left(t_{1},t_{3}\right)\mathbf{G}\left(\mathbf{x_{1}}+\intop_{t_{3}}^{t_{1}}\mathbf{v}\left(t'\right)dt',\mathbf{x_{2}},t_{1},t_{2}\right)\\=\mathbb{1}_{3}\delta^{(3)}\left(\mathbf{x_{1}}-\mathbf{x_{2}}\right)\delta\left(t_{1}-t_{2}\right).
\end{split}
\end{equation}
where \(\chi(t,t')\) is equal to (\ref{movsuc}) without the spatial delta function.  Finding the solution of (\ref{eq:GreensFunctionWaveEquation}) would be essential in any calculation involving the electric field.  In the cases treated in \cite{TomMacro}, the difficulty of finding solutions resides in the complexity of the geometry. In this case though, even in homogeneous media, the integro-differential equation above presents further difficulties due to its non-locality.  Yet in a few simple cases, some progress can be made.  Given the homogeneity of the medium in this case, (\ref{eq:GreensFunctionWaveEquation}) can by Fourier transformed so that it reduces to the integral equation
\begin{equation}\label{eq:FouierWaveEq}
	\mathbf{k}\times\mathbf{k}\times\mathbf{G}\left(\mathbf{k},\Omega,\Omega_{2}\right)+\frac{\Omega^{2}}{c^{2}}\mathbf{G}\left(\mathbf{k},\Omega,\Omega_{2}\right)+\frac{\Omega^{2}}{c^{2}}\intop_{-\infty}^{\infty}\frac{d\Omega_{1}}{2\pi}\chi\left(\mathbf{k},\Omega,\Omega_{1}\right)\mathbf{G}\left(\mathbf{k},\Omega_{1},\Omega_{2}\right)=-2\pi\mathbb{1}_{3}\delta(\Omega-\Omega_{2})
	\end{equation} 
where the kernel of the integral represents the Fourier transformed susceptibility of the system and is given by
\begin{equation}\label{eq:FourierSusc}
	\chi\left(\mathbf{k},\Omega,\Omega_{1}\right)=\intop_{-\infty}^{\infty}dt\intop_{-\infty}^{t}dt_{1}\ \chi\left(t,t_{1}\right)e^{i\mathbf{k}.\intop_{t_{1}}^{t}\mathbf{v}\left(t'\right)dt'}e^{i\Omega t}e^{-i\Omega_{1}t_{1}}.
\end{equation} 
For the case of constant velocity \(\mathbf{v}(t)=\mathbf{v}\), the wave equation (\ref{eq:GreensFunctionWaveEquation}) simplifies to that given in~\cite{MovingMedia} (again, ignoring polarisation mixing) where the Doppler shifted frequency \(\Omega-\mathbf{v}\cdot\mathbf{k}\) appears in the argument of the susceptibility,
\begin{equation}\label{eq:doppler}
	\left\{\mathbf{k}\otimes\mathbf{k}-\left[k^{2}-\frac{\Omega^{2}}{c^{2}}\left(1+\chi\left(\Omega-\mathbf{k\cdot v}\right)\right)\right]\mathbb{1}_{3}\right\}\cdot\mathbf{G}\left(\mathbf{k},\Omega,\Omega_{2}\right)=-2\pi\delta(\Omega-\Omega_{2})\mathbb{1}_{3}.
\end{equation}
The Green function is then equal to the inverse of the square bracketed matrix on the left times \(2\pi\mu_{0}\delta(\Omega-\Omega_{2})\).  In other words, the constant motion of the dielectric, gives rise to a new effective permittivity in which the frequency response is shifted by $-\mathbf{k}\cdot\mathbf{v}$. 
\par
Now consider a slightly more complicated case, that of a time dependent oscillatory motion in the $z$-direction, with frequency $\nu$, \(\mathbf{v}(t)=z_{0}\nu\sin(\nu t)\hat{\mathbf{z}}\), where $z_{0}$ is the maximum displacement. In this case the integral over the velocity in the susceptibility (\ref{eq:FourierSusc}) becomes
\begin{equation}
	\mathbf{k}.\intop_{t_{1}}^{t}\mathbf{v}\left(t'\right)dt'=k_{z}z_{0}\cos(\nu t_{1})-k_{z}z_{0}\cos(\nu t)
\end{equation}
The susceptibility (\ref{eq:FourierSusc}) can then be written as
\begin{equation}
	\chi\left(\mathbf{k},\Omega,\Omega_{1}\right)=2\pi\sum_{n=-\infty}^{\infty}\sum_{m=-\infty}^{\infty}i^{n-m}J_{n}(k_{z}z_{0})J_{m}(k_{z}z_{0})\delta(\Omega-\Omega_{1}+(m+n)\nu)\chi(\Omega_{1}-n\nu).\label{oscsuc}
\end{equation}
where we applied the generating function for Bessel functions~\cite{MathMeth}
\begin{equation}
e^{iz\cos\theta}=\sum_{n=-\infty}^{\infty}i^{n}J_{n}\left(z\right)e^{in\theta}
\end{equation}
where $J_{n}$ are the Bessel functions of the first kind.
Expression (\ref{oscsuc}) demonstrates a coupling between the frequencies of the field arising from the oscillation, occurring in discrete multiples of the oscillation frequency \(\nu\). It is interesting to note that the susceptibility vanishes when $k_{z}z_{0}$ equals a zero of a Bessel function, which for a fixed \(\mathbf{k}\) and \(\Omega\) excludes the coupling to those frequencies \(\Omega_{1}-(n+m)\nu\) where \(J_{n,m}(k_{z}z_{0})=0\).  In the limiting case when the argument of the Bessel functions, $k_{z}z_{0}$, becomes very large, the material behaves like the vacuum since the Bessel functions all tend to zero. For very small $k_{x}z_{0}$, only the $m=n=0$ survives and the material behaves as if it were stationary. This occurs for very large wavelengths or in the limit of a vanishing displacement amplitude.  It is still difficult to exactly solve (\ref{eq:FouierWaveEq}) for the Green function \(\mathbf{G}\left(\mathbf{k},\Omega,\Omega_{2}\right)\).  However, if we make for example the simplifying approximation that \(\nu\ll\Omega\) and \(k_{z}z_{0}\sim 1\) or smaller then the delta function and the susceptibility can be removed from under the summation sign in (\ref{oscsuc}) and the system behaves as if it had the effective permittivity given by the sum over the product of the Bessel functions.  
\par
The above brief examination of the classical physics of electromagnetism interacting with a moving dielectrc demonstrates that for most cases of interest it is difficult to find exact solutions to the equations of motion.  In order to establish quantitative results in general, proceeding via perturbation theory seems the best approach and is the one we adopt from this point on. 

\section{Emission rates for time-dependent media}
\par
We now quantize the classical theory outlined in the previous section, applying it to the situations illustrated in figure \ref{fig:System}.  Most of the formula remain formally quite similar, but we must replace the classical fields with field operators that obey the canonical commutation relations \cite{QED}, 
\begin{equation}
\begin{split}
	&\left[\mathbf{\hat{A}}\left(\mathbf{x},t\right),\mathbf{\hat{\Pi}}_{\mathbf{\mathbf{A}}}\left(\mathbf{x'},t\right)\right]=i\hbar\mathbb{1}_{3}\delta\left(\mathbf{x}-\mathbf{x'}\right)
 \\&\left[\mathbf{\hat{X}}_{\omega}\left(\mathbf{x},t\right),\mathbf{\hat{\Pi}}_{\mathbf{X}_{\omega'}}\left(\mathbf{x'},t\right)\right]=i\hbar\mathbb{1}_{3}\delta\left(\mathbf{x}-\mathbf{x'}\right)\delta\left(\omega-\omega'\right).
\end{split}
\end{equation}
We work in the interaction picture~\cite{QED}, where the time dependence of the operators is generated by the bare Hamiltonian $\hat{H}_{0}$, given by the operator equivalent of (\ref{eq:H0}) and that of the quantum state by the interaction Hamiltonian $\hat{H}_{I} (t) = e^{-i\hat{H}_{0}t}\hat{H}_{I}e^{i\hat{H}_{0}t}$ given by the operator equivalent of (\ref{eq:Pert}) where the full Hamiltonian is given by $\hat{H}=\hat{H}_{0}+\hat{H}_{I}$.  
\par
With the time dependence of the operators being generated by \(\hat{H}_{0}\), the canonical operators are expanded in terms of a set of creation and annihilation operators $\mathbf{\hat{C}}\left(\mathbf{x},\omega\right)$, $\mathbf{\hat{C}}^{\dagger}\left(\mathbf{x},\omega\right)$ as
\begin{align}\label{expansion}
	\mathbf{\hat{A}}\left(\mathbf{x},t\right)&=\int d^{3}\mathbf{x}'\intop_{0}^{\infty}d\omega'\left[ \mathbf{f}_{\mathbf{A}}\left(\mathbf{x},\mathbf{x'},\omega'\right)\cdot\mathbf{\hat{C}}\left(\mathbf{x'},\omega'\right)e^{-i\omega't}+h.c\right]\\
	\hat{\Pi}_{\mathbf{A}}\left(\mathbf{x},t\right)&=\int d^{3}\mathbf{x}'\intop_{0}^{\infty}d\omega'\left[ \mathbf{f}_{\mathbf{\Pi_{A}}}\left(\mathbf{x},\mathbf{x'},\omega'\right)\cdot\mathbf{\hat{C}}\left(\mathbf{x'},\omega'\right)e^{-i\omega't}+h.c\right]\\
	\mathbf{\hat{X}}_{\omega}\left(\mathbf{x},t\right)&=\int d^{3}\mathbf{x}'\intop_{0}^{\infty}d\omega'\left[ \mathbf{f}_{\mathbf{X}}\left(\mathbf{x},\mathbf{x'},\omega,\omega'\right)\cdot\mathbf{\hat{C}}\left(\mathbf{x'},\omega'\right)e^{-i\omega't}+h.c\right]\\
	\hat{\Pi}_{\mathbf{X_{\omega}}}\left(\mathbf{x},t\right)&=\int d^{3}\mathbf{x}'\intop_{0}^{\infty}d\omega'\left[ \mathbf{f}_{\mathbf{\Pi_{X}}}\left(\mathbf{x},\mathbf{x'},\omega,\omega'\right)\cdot\mathbf{\hat{C}}\left(\mathbf{x'},\omega'\right)e^{-i\omega't}+h.c\right]
\end{align}
where the expansion coefficients are identical to those given in \cite{TomMacro}, and are given for reference here in appendix~\ref{ApA}. There are also similar expansions for the $\hat{\mathbf{E}}$ and $\hat{\mathbf{B}}$ operators.  The operators $\mathbf{\hat{C}}$ and $\mathbf{\hat{C}}^{\dagger}$ represent the creation and annihilation of quanta of the coupled field--matter system (polaritons) and obey bosonic commutation relations,
\begin{equation}\label{commutation}
\begin{split}
	&\left[\mathbf{\hat{C}}\left(\mathbf{x},\omega\right),\mathbf{\hat{C}}^{\dagger}\left(\mathbf{x'},\omega'\right)\right]=\mathbb{1}_{3}\delta\left(\mathbf{x}-\mathbf{x'}\right)\delta\left(\omega-\omega'\right)
 \\&\left[\mathbf{\hat{C}}\left(\mathbf{x},\omega\right),\mathbf{\hat{C}}\left(\mathbf{x'},\omega'\right)\right]=0
\end{split}
\end{equation}
The expansion given in (\ref{expansion}) is chosen such that it diagonalises the bare Hamiltonian, 
\begin{equation}
	\hat{H}_{0}=\intop d^{3}\mathbf{x}\intop_{0}^{\infty}d\omega\ \hbar\omega\ \mathbf{\hat{C}}^{\dagger}\left(\mathbf{x},\omega\right)\mathbf{\hat{C}}\left(\mathbf{x},\omega\right)
\end{equation}
For more details about this diagonalization see~\cite{HuttnerBarnettModel,Suttorp2004,ScheelQED,TomMacro,MovingMedia}.
\par
We now apply the usual methods of time dependent perturbation theory to find the effect of the motion or time dependence of the medium on the excitation of polaritons.  In the absence of any perturbation, we take the system to be prepared in its ground state $\left|0\right\rangle$ (defined as the state where \(\hat{\boldsymbol{C}}_{\omega}|0\rangle=0\)).  Given that, to leading order, the introduction of the interaction Hamiltonian \(\hat{H_{I}}\) will lead to the creation of pairs of polaritons we represent the wave function of the system as
\begin{equation}
	\left|\psi(t)\right\rangle=\left| 0\right\rangle+\sum_{m,n}\ \intop d^{3}\mathbf{x}_{1}\intop d^{3}\mathbf{x}_{2}\intop_{0}^{\infty}d\omega_{1}\intop_{0}^{\infty}d\omega_{2}\ 
\zeta_{mn}\left(\mathbf{x_{1}},\mathbf{x_{2}},\omega_{1},\omega_{2},t\right)\hat{C}_{m}^{\dagger}\left(\mathbf{x}_{1},\omega_{1}\right)\hat{C}_{n}^{\dagger}\left(\mathbf{x}_{2},\omega_{2}\right)\left|0\right\rangle\label{psi}
\end{equation}
where the expansion coefficient obeys \(\zeta_{mn}\left(\mathbf{x_{1}},\mathbf{x_{2}},\omega_{1},\omega_{2},t\right)=\zeta_{nm}\left(\mathbf{x_{2}},\mathbf{x_{1}},\omega_{2},\omega_{1},t\right)\), in accordance with bosonic exchange symmetry.  The physical meaning of \(\zeta\) is as the probability amplitude for the creation of a pair of current excitations in the material, located at positions \(\mathbf{x}_{1}\) and \(\mathbf{x}_{2}\), oscillating at frequencies \(\omega_{1}\) and \(\omega_{2}\), and pointing in the \(m\) and \(n\) directions.  Inserting (\ref{psi}) into the Sch\'odinger equation \(\partial |\psi\rangle/\partial t=-(i/\hbar)\hat{H}_{I}|\psi\rangle\), we find the rate of change of the expansion coefficient is given by
\begin{equation}\label{asymZeta}
	\dot{\zeta}_{mn}\left(\mathbf{x_{1}},\mathbf{x_{2}},\omega_{1},\omega_{2},t\right)=-\frac{i}{2\hbar}\langle 0|\hat{C}_{m}(\mathbf{x}_{1},\omega_{1})\hat{C}_{n}(\mathbf{x}_{2},\omega_{2})\hat{H}_{I}|0\rangle
\end{equation}
Due to the form of the interaction Hamiltonian (\ref{eq:Pert}), the rate of change of the rank-2 tensor $\zeta_{mn}$ separates into the sum of two parts,
\begin{equation}\label{rate}
	\dot{\zeta}_{mn}\left(\mathbf{x_{1}},\mathbf{x_{2}},\omega_{1},\omega_{2},t\right)=
\dot{\zeta}_{mn}^{(a)}\left(\mathbf{x_{1}},\mathbf{x_{2}},\omega_{1},\omega_{2},t\right)
+\dot{\zeta}_{mn}^{(b)}\left(\mathbf{x_{1}},\mathbf{x_{2}},\omega_{1},\omega_{2},t\right).
\end{equation}
The first term $\zeta_{mn}^{(a)}$ arises from the first two terms in the interaction Hamiltonian (\ref{eq:Pert}) and represents the effect of the motion of the medium, both through the movement of energy within the bath of oscillators, and the mixing of the electromagnetic field polarization between reference frames. The second term, $\zeta_{mn}^{(b)}$ comes from the final term in (\ref{eq:Pert}), and represents the effect of any time dependence in the permittivity.  Evaluating (\ref{asymZeta}) by substituting the expansion of the operators (\ref{expansion}) into the interaction Hamiltonian and using the commutation relations (\ref{commutation}), the dyadic form the rate of change of the two parts to the expansion coefficient (\ref{rate}) is found to be
\begin{equation}\label{zeta_v}
\begin{split}
	\dot{\boldsymbol{\zeta}}^{(a)}\left(\mathbf{x_{1}},\mathbf{x_{2}},\omega_{1},\omega_{2},t\right)&=-\frac{i}{2\hbar}e^{i(\omega_{1}+\omega_{2})t}\int d^{3}\mathbf{x}\intop_{0}^{\infty}d\omega\ \Bigg[\mathbf{f}_{\mathbf{\Pi_{X}}}^{\dagger}\left(\mathbf{x},\mathbf{x_{1}},\omega,\omega_{1}\right)\cdot\left(\mathbf{v}\cdot\nabla\right)\mathbf{f}_{\mathbf{X}}^{*}\left(\mathbf{x},\mathbf{x_{2}},\omega,\omega_{2}\right)\\&+\mathbf{f}_{\mathbf{B}}^{\dagger}\left(\mathbf{x},\mathbf{x_{1}},\omega_{1}\right)\cdot\boldsymbol{\alpha}_{B}\left(\omega,\mathbf{x},t\right)\cdot \mathbf{f}_{\mathbf{X}}^{*}\left(\mathbf{x},\mathbf{x_{2}},\omega,\omega_{2}\right)\Bigg]+1\leftrightarrow2
\end{split}
\end{equation}
and
\begin{equation}\label{secondTerm}
	\dot{\boldsymbol{\zeta}}^{(b)}\left(\mathbf{x_{1}},\mathbf{x_{2}},\omega_{1},\omega_{2},t\right)=-\frac{i}{2\hbar}e^{i(\omega_{1}+\omega_{2})t}\int d^{3}\mathbf{x}\intop_{0}^{\infty}d\omega\ \delta\alpha\left(\omega,\mathbf{x},t\right)\mathbf{f}_{\mathbf{E}}^{\dagger}\left(\mathbf{x},\mathbf{x_{1}},\omega_{1}\right)\cdot \mathbf{f}_{\mathbf{X}}^{*}\left(\mathbf{x},\mathbf{x_{2}},\omega,\omega_{2}\right)+1\leftrightarrow2
\end{equation} 
where the notation `\(1\leftrightarrow2\)' indicates the repetition of the preceding expression but with the two particles interchanged (which in the above expressions involves both taking the transpose and swapping subscripts \(1\) and \(2\)).  It is evident from expressions (\ref{zeta_v}) and (\ref{secondTerm}) that the rate of polariton excitation is in general quite different for time dependent and moving media.  For instance, a time dependent permittivity profile constructed to appear as a moving material would have \(\boldsymbol{\zeta}^{(a)}=0\), whereas true motion has in general both non--zero \(\boldsymbol{\zeta}^{(a)}\) and \(\boldsymbol{\zeta}^{(b)}\). 
\par
In the remainder of the paper we shall evaluate polariton excitation rates implied by (\ref{zeta_v}) and (\ref{secondTerm}) for the three cases shown in figure~\ref{fig:System}.  To be specific we assume the simplest case of a lossy dispersive medium, where the background dielectric function \(\epsilon_{b}\) is Lorentzian, with a resonant frequency $\omega_{0}$
\begin{equation}\label{lorentzian}
	\epsilon_{b}(\omega)=1+\frac{\omega^{2}_{p}}{\omega_{0}^2-\omega^2-i\gamma\omega}
\end{equation}
where we assume the arbitrary values $\omega_{p}=0.5\ \omega_{0}$, and damping constant $\gamma=0.1\ \omega_{0}$.
\section{Emission due to the motion of a homogeneous medium\label{homed}}
\par
The first calculation we perform is the polariton emission rate within an infinite homogeneous medium performing an oscillatory motion.  Due to the lack of boundaries and the translational invariance, the second term in (\ref{rate}), $\dot{\boldsymbol{\zeta}}^{(b)}$ is equal to zero, and the probability amplitude for emitting a pair of polaritons is then equal to the integral of \(\dot{\boldsymbol{\zeta}}^{(a)}\) over the time interval \(t\in[-T/2,T/2]\) (as is typical in such calculations, it is assumed that the interaction Hamiltonian is `turned on' at the initial time \(-T/2\)).  The absolute value squared of the result divided by $T$ gives us the average net excitation rate of polariton pairs over the time interval $T$.  Over a very long time interval \(T\to\infty\) this is given by
\begin{equation}\label{eq:Fermi}
	\Gamma=\intop d^{3}\mathbf{x_{1}}\intop d^{3}\mathbf{x_{2}}\intop_{0}^{\infty}d\omega_{1}\intop_{0}^{\infty}d\omega_{2}\ \sum_{m,n}\underset{T\rightarrow\infty}{\lim}\ \frac{1}{T}\left|\intop_{-T/2}^{T/2}dt\ \dot{\zeta}^{(a)}_{mn}\left(\mathbf{x_{1}},\mathbf{x_{2}},\omega_{1},\omega_{2},t\right)\right|^{2}.
\end{equation}
The emission rate given by (\ref{eq:Fermi}) is valid even for spatially dependant velocities \(\mathbf{v}\), as occurs e.g. for rotating bodies.  However here we make the simplification that the velocity does not depend on position, in which case---as is evident from (\ref{zeta_v})---the time dependence of \(\zeta^{(a)}\) can be factored out from the spatial dependence.  For example, in the case of constant velocity the time dependence of (\ref{zeta_v}) is given by the factor \(\exp(i(\omega_{1}+\omega_{2})t)\) so that
\begin{equation}\label{eq:timedep}
	\underset{T\rightarrow\infty}{\lim}\ \frac{1}{T}\left|\intop_{-T/2}^{T/2}dt\ e^{i\left(\omega_{1}+\omega_{2}\right)t}\right|^{2}=2\pi\delta\left(\omega_{1}+\omega_{2}\right).
\end{equation}
The emission rate (\ref{eq:Fermi}) is then reduced to
\begin{equation}\label{emissionexp}
	\Gamma=2\pi\intop d^{3}\mathbf{x_{1}}\intop d^{3}\mathbf{x_{2}}\intop_{0}^{\infty}d\omega_{1}\intop_{0}^{\infty}d\omega_{2}\ \sum_{m,n}\left|\zeta^{(a)}_{mn}\left(\mathbf{x_{1}},\mathbf{x_{2}},\omega_{1},\omega_{2}\right)\right|^{2}\delta\left(\omega_{1}+\omega_{2}\right)
\end{equation} 
where $\zeta^{(a)}_{\mu \nu}\left(\mathbf{x_{1}},\mathbf{x_{2}},\omega_{1},\omega_{2},t\right)=\zeta^{(a)}_{\mu \nu}\left(\mathbf{x_{1}},\mathbf{x_{2}},\omega_{1},\omega_{2}\right)\exp\left(i\left(\omega_{1}+\omega_{2}\right)t\right)$. The argument of the delta function is only zero at the one point \(\omega_{1}=\omega_{2}=0\), where the integrand is zero.  This leaves,
\begin{equation}
	\Gamma=0
\end{equation} 
as expected.
 \par
When the velocity exhibits a time dependence then in general the emission rate \(\Gamma\) will be non--zero.  Consider some time dependent velocity $\mathbf{v}(t)=v(t)\hat{\mathbf{z}}$.  In order to analyse the emission rate as a function of frequency and wave--vector, we work in terms of the Fourier transform of $\boldsymbol{\zeta}^{(a)}$,
\begin{equation}\label{defA}
\begin{split}
	\dot{\boldsymbol{\zeta}}^{(a)}\left(\mathbf{x_{1}},\mathbf{x_{2}},\omega_{1},\omega_{2},t\right)&=
\intop\frac{d^{3}\mathbf{k}}{\left(2\pi\right)^{3}}\dot{\boldsymbol{\zeta}}^{(a)}\left(\mathbf{k},\omega_{1},\omega_{2},t\right)e^{i\mathbf{k}.\left(\mathbf{x_{1}}-\mathbf{x_{2}}\right)}
\\&=v\left(t\right)e^{i\left(\omega_{1}+\omega_{2}\right)t}\intop\frac{d^{3}\mathbf{k}}{\left(2\pi\right)^{3}}\boldsymbol{\mathcal{A}}\left(\mathbf{k},\omega_{1},\omega_{2}\right)e^{i\mathbf{k}.\left(\mathbf{x_{1}}-\mathbf{x_{2}}\right)}
\end{split}
\end{equation} 
%
where the quantity $\boldsymbol{\mathcal{A}}\left(\mathbf{k},\omega_{1},\omega_{2}\right)$ is given in appendix~\ref{ApB} and is a sum of terms depending of products of Green's functions and combinations of the permittivity at frequencies $\omega_{1}$ and $\omega_{2}$. The amplitude $\boldsymbol{\mathcal{A}}(\mathbf{k},\omega_{1},\omega_{2})$ is proportional to the probability amplitude for exciting a pair of polaritons with wave-vectors $\mathbf{k}$ and $\mathbf{-k}$ and frequencies $\omega_{1}$ and $\omega_{2}$ from the ground state, due to the motion of the medium.
\par
Consider the specific case of an oscillatory motion with frequency $\nu$: $v(t)=z_{0}\nu\cos(\nu t)$, where $z_{0}$ is the maximum displacement from the mean position.   Taking the absolute value squared of (\ref{defA}) and integrating with respect to time, the equivalent of (\ref{eq:timedep}) now equals
\begin{equation}\label{eq:delta}
	\underset{T\rightarrow\infty}{\lim}\ \frac{1}{T}\left|\intop_{-T/2}^{T/2}dt\ v\left(t\right)e^{i\left(\omega_{1}+\omega_{2}\right)t}\right|^{2}=\frac{\pi z_{0}^{2}\nu^{2}}{2}\left[\delta\left(\omega_{1}+\omega_{2}-\nu\right)+\delta\left(\omega_{1}+\omega_{2}+\nu\right)\right]
\end{equation}
the proof of which is given in appendix.  While the second term on the right of (\ref{eq:delta}) fails to contribute to the emission rate (owing to $\omega_{1}$, $\omega_{2}$ and $\nu$ all being positive), the first term does.  Combining (\ref{defA}) and (\ref{eq:delta}) with the expression for the net rate of excitation (\ref{emissionexp}) gives the emission rate per unit volume \(V\)
\begin{equation}\label{eq:excitation}
	\frac{\Gamma}{V}=\int\frac{d^{3}\mathbf{k}}{(2\pi)^{3}}\intop_{0}^{\infty} d\omega\ \rho(\mathbf{k},\omega)
\end{equation}
where the `spectral density' for the emission of polariton pairs \(\rho\) is equal to
\begin{equation}\label{eq:emissdens}
	\rho(\mathbf{k},\omega)=\frac{\pi z_{0}^{2}\nu^{2}}{2}\theta(\nu-\omega)\sum_{m,n}\left|\mathcal{A}_{mn}(\mathbf{k},\omega,\nu-\omega)\right|^{2}
\end{equation}
with $\theta$ the Heaviside step function.  This dimensionless spectral density depends only on the wave vector and frequency since the translational symmetry of this system guarantees momentum conservation for the two polaritons, \(\mathbf{k}_{1}+\mathbf{k}_{2}=0\), and energy conservation implies that the total energy of the pair of particles must equal that due to the motion: \(\hbar\nu=\hbar(\omega_{1}+\omega_{2})\).  Note that the time dependence of the velocity sets this relationship between $\omega_{1}$ and $\omega_{2}$,  and that for a general motion this may be more complicated.
%
%
%
\par
The excitation rate (\ref{eq:excitation}) evidently scales quadratically with the maximum displacement $z_{0}$. However, the dependence on the oscillation frequency \(\nu\) is somewhat more intricate.  As can be seen from the expression for $\boldsymbol{\mathcal{A}}$ given in appendix~\ref{ApB}, as well as figure \ref{fig:EmissionAmplitude_Full}a, the largest contribution to the spectral density \(\rho\) comes from regions where the frequency and wave vector are close to satisfying the dispersion relations for electromagnetic waves \(\omega=c|\mathbf{k}|/\epsilon^{\star}(\omega)\) and \(\nu-\omega=\pm c|\mathbf{k}|/\epsilon^{\star}(\nu-\omega)\) and where the frequency matches the resonant frequency of the dielectric \(\omega_{0}\), and the shifted resonance at \(\nu-\omega_{0}\).  The largest emission occurs in the region close to where the dispersion curves for radiation intersect with the material resonances.  
\par
We note that there is a term that contributes to (\ref{eq:excitation})---given as $d(\omega,\nu-\omega)$ in appendix~\label{ApB}---which diverges when $\omega_{1}=\nu/2$ but is zero for all other frequencies. This has been omitted from figure \ref{fig:EmissionAmplitude_Full}a.  Physically, this problematic term represents the emission rate for two polaritons at the same frequency from the same position.  This is specific to moving media and does not recur in the remainder of the paper.  It nevertheless deserves further attention and will be treated in future work.
%
%
\begin{figure}[h]
		\includegraphics[height=7cm]{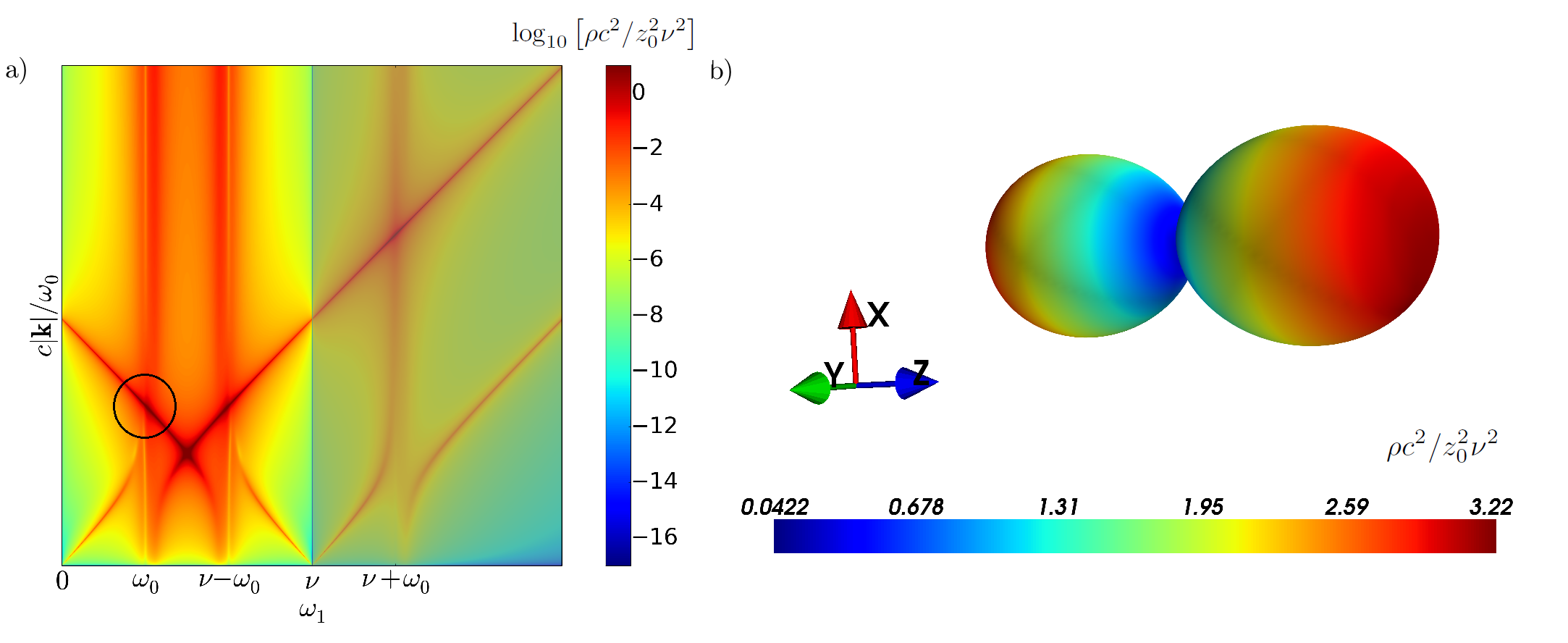}
		\caption{a): Logarithm of the spectral density of polariton emission $\log_{10}\left[\rho c^{2}/z^{2}_{0}\nu^{2}\right]$, due to the oscillatory motion of a homogeneous dielectric.  The dielectric function of the material \(\epsilon_{b}\) is given by the Lorentzian response (\ref{lorentzian}) with a resonance at $\omega_{0}$, while the oscillation frequency of the motion is arbitrarily chosen as $\nu=3\omega_{0}$ to clearly separate the curves.  The spectral density is mostly concentrated around frequencies in the vicinity of the resonant frequency of the material $\omega_{0}$, the shifted resonant frequency $\nu-\omega_{0}$ as well as along the two dispersion curves satisfying $c^{2}k^{2}-\epsilon^{*}(\omega)
\omega^{2}=0$, and the shifted dispersion curve $c^{2}k^{2}-\epsilon^{*}(\nu-\omega)(\nu-\omega)^{2}=0$. The vertical line at $\omega=\nu$ represents the upper boundary of the area contributing to the emission rate (\ref{eq:excitation}); the shaded area to right of this does not contribute to the total emission rate. The black circle points to an example of a region where the shifted dispersion curve intersects with the resonance at \(\omega_{0}\); the emission rate density is much more intense and somewhat spread out around these areas. b): Angular dependence of spectral density $\rho(\mathbf{k})c^{2}/z^{2}_{0}\nu^{2}$ where the frequency and wave vector magnitude are arbitrarily chosen to be $\omega=\omega_{0}$ and $|\mathbf{k}|=2\omega_{0}$ respectively corresponding to the centre of the black circle in (a). The distance of the surface from the centre of the graph shows the amplitude of the emission rate for varying directions of $\mathbf{k}$. The strongest excitation of polaritons occurs in the direction of motion.}
		\label{fig:EmissionAmplitude_Full}
\end{figure}

\section{Time Dependent Permittivity}

The second contribution to the emission rate (\ref{secondTerm}) \(\dot{\boldsymbol{\zeta}}^{(b)}\) comes from the time dependence of the material response $\delta\alpha\left(\mathbf{x},t,\omega\right)$.  For moving media, this term can be non--zero either due to moving inhomogeneities or changing boundaries.  This term can also be used to model changes in the permittivity due to external forces, such as those in dynamical Casimir experiments \cite{DynCas,ObservationDynCas} or optical analogues of Hawking radiation \cite{AnalogHawking}, examples of both of which we now consider.
\par
Inserting the expansion coefficients listed in appendix~\ref{ApA} into (\ref{secondTerm}), we find the rate of change of the probability amplitude for exciting a pair of polaritons is given by
\begin{equation}\label{zetadyn}
\begin{split}
	\dot{\boldsymbol{\zeta}}^{(b)}\left(\mathbf{x_{1}},\mathbf{x_{2}},\omega_{1},\omega_{2},t\right)=&-\frac{i\mu_{0}\omega_{1}^{2}\alpha_{b}\left(\omega_{1}\right)}{4\sqrt{\omega_{1}\omega_{2}}}e^{i\left(\omega_{1}+\omega_{2}\right)t}\times\\&\Bigg[\mu_{0}\omega_{2}^{2}\alpha_{b}\left(\omega_{2}\right)\intop_{0}^{\infty}d\omega\frac{\alpha_{b}\left(\omega\right)}{\omega{}^{2}-\left(\omega_{2}-i\omega0^{+}\right)^{2}}\int d^{3}\mathbf{x}\ \delta\alpha\left(\omega,\mathbf{x},t\right)\ \mathbf{G}^{\dagger}\left(\mathbf{x},\mathbf{x_{1}},\omega_{1}\right)\cdot\mathbf{G}^{*}\left(\mathbf{x},\mathbf{x_{2}},\omega_{2}\right)\\&
+\delta\alpha\left(\omega_{2},\mathbf{x_{2}},t\right)\mathbf{G}^{\dagger}\left(\mathbf{x_{2}},\mathbf{x_{1}},\omega_{1}\right)\Bigg]
+1\leftrightarrow2
\end{split}
\end{equation}
We now consider two particular applications of (\ref{zetadyn}), firstly where the permittivity of the medium uniformly oscillates as a function of time, and secondly where a travelling pulse-like perturbation $\delta\alpha\left(\mathbf{x}-vt,\omega\right)$ to the permittivity moves through the medium at a uniform velocity. 

\subsection{Emission from a Time Dependent Permittivity\label{sec:tdp}}
\par
\begin{figure}[h]
		\includegraphics[height=7cm]{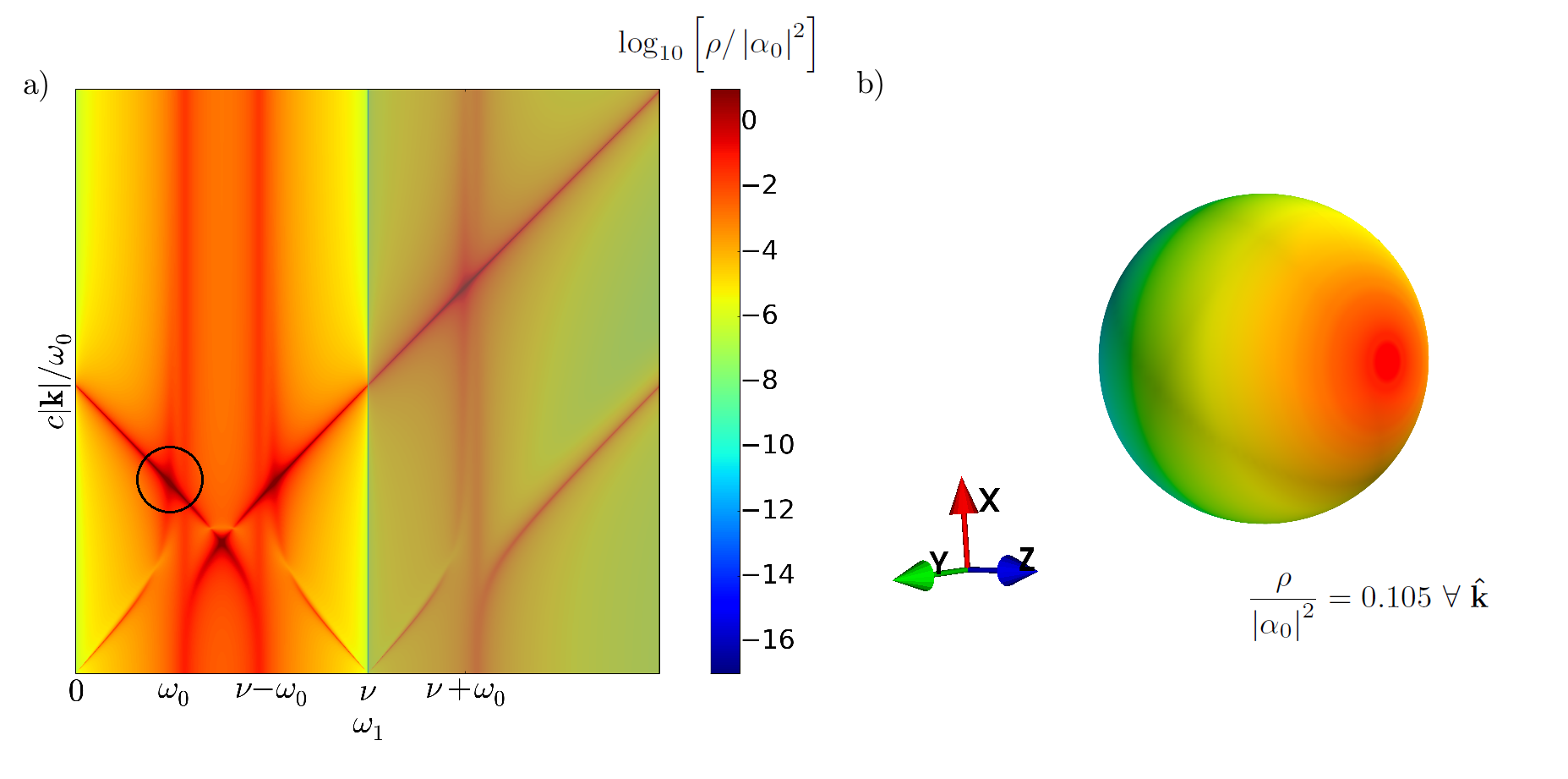}
		\caption{a):  Logarithm of the spectral density of polariton emission $\log_{10}\left[\rho/|\alpha_{0}|^{2}\right]$ as a function of $\omega$ and $c|\mathbf{k}|$, for a dielectric function oscillating as a function of time with frequency \(\nu\).  The background dielectric function of the material \(\epsilon_{b}\) is given by the Lorentzian response (\ref{lorentzian}) with a resonance at $\omega_{0}$, while the oscillation frequency of the material properties is chosen as $\nu=3\omega_{0}$ to clearly separate the curves. The regions of high emission occur at the same points as shown for the moving dielectric in figure~\ref{fig:EmissionAmplitude_Full}.  b): Angular dependence of spectral density $\rho(\mathbf{k})/|\alpha_{0}|^{2}$ where the frequency and wave vector magnitude are arbitrarily chosen to be $\omega=\omega_{0}$ and $|\mathbf{k}|=2\omega_{0}$ respectively corresponding to the centre of the black circle in (a). The colour in this subfigure only serves to illustrate that the shape is a sphere, indicating isotropic emission.}
		\label{fig:DynamicAmplitude}
\end{figure}

Consider a time dependent change to the permittivity of the form $\delta\mathbf{\alpha}\left(\omega,t\right)=\alpha_{0}\beta\left(\omega\right)\cos\left(\nu t\right)$ corresponding to a periodic change in the medium over all space with frequency $\nu$. In this case, the time dependence can be factored out as in the previous section, and the identity (\ref{eq:delta}) can be applied.  The spectral density appearing in the polariton emission rate per unit volume (the analogue of (\ref{eq:excitation})) is again dimensionless and found to be
\begin{equation}\label{eq:EmissDensPerm}
\begin{split}
\rho\left(\mathbf{k},\omega\right)&=\frac{\pi}{2}\intop_{0}^{\infty}d\omega_{2}\ \delta\left(\omega+\omega_{2}-\nu\right)\sum_{m,n}\left|\mathcal{A}_{mn}\left(\mathbf{k},\omega,\omega_{2}\right)\right|^{2}
\\&=\frac{\pi}{2}\sum_{m,n}\left|\mathcal{A}_{mn}\left(\mathbf{k},\omega,\nu-\omega\right)\right|^{2}\theta(\nu-\omega)
\end{split}
\end{equation}
where
%
%
%
\begin{equation}\label{eq:DynCasCoeffGen}
\begin{split}
	\boldsymbol{\mathcal{A}}\left(\mathbf{k},\omega_{1},\omega_{2}\right)&=-\frac{i\mu_{0}\alpha_{0}\omega_{1}^{2}\alpha_{b}(\omega_{1})}{4\sqrt{\omega_{1}\omega_{2}}}\Bigg[\mu_{0}\alpha_{b}\left(\omega_{2}\right)\omega_{2}^{2}\intop_{0}^{\infty}d\omega\frac{\alpha_{b}\left(\omega\right)\beta\left(\omega\right)}{\omega{}^{2}-\left(\omega_{2}-i\omega0^{+}\right)^{2}}\mathbf{G}^{\dagger}\left(\mathbf{k},\omega_{1}\right)\cdot\mathbf{G}^{*}\left(-\mathbf{k},\omega_{2}\right)+\beta\left(\omega_{2}\right)\mathbf{G}^{\dagger}\left(\mathbf{k},\omega_{1}\right)\Bigg]\\&+1\leftrightarrow2.
\end{split}
\end{equation}
with `\(1\leftrightarrow2\)' again indicating an interchange of the two particles (which now involves swapping the subscripts \(1\) and \(2\), interchanging \(\mathbf{k}\) for \(-\mathbf{k}\) and taking the transpose).  As in the case of the moving dielectric that we just discussed, the emission conserves momentum \(\mathbf{k}_{1}+\mathbf{k}_{2}=0\) and the energy of the pair of polaritons is taken from the oscillation: \(\hbar\nu=\hbar(\omega_{1}+\omega_{2})\).  Equation (\ref{eq:DynCasCoeffGen}) can be further simplified in the particular case where $\beta\left(\omega\right)=\alpha_{b}\left(\omega\right)$ (i.e. the change in the permittivity has the same frequency dependent response as the background), 
\begin{equation}
	\boldsymbol{\mathcal{A}}\left(\mathbf{k},\omega_{1},\omega_{2}\right)=-\frac{i\alpha_{0}}{2\pi}\sqrt{{\rm Im}[\chi_{b}\left(\omega_{1}\right)]{\rm Im}[\chi_{b}\left(\omega_{2}\right)]}\frac{\omega^{2}_{1}}{c^{2}}\mathbf{G}^{\dagger}\left(\mathbf{k},\omega_{1}\right)\cdot\left[ \chi_{b}^{\star}\left(\omega_{2}\right)\frac{\omega^{2}_{2}}{c^{2}}\mathbf{G}^{*}\left(-\mathbf{k},\omega_{2}\right)+\mathbb{1}_{3}\right]+1\leftrightarrow2.
\end{equation}
As is evident from (\ref{eq:EmissDensPerm})---and in similarity to the time dependent material velocity investigated in the previous section---the time dependence of $\delta\alpha$ gives rise to emission only for frequencies in the range $\omega_{1}>\nu$.  The emission rate per unit volume is then given by integrating (\ref{eq:EmissDensPerm}) over wave vector and frequency.
%
The frequency and wave vector dependence of the spectral density of polariton emission are plotted in Figure \ref{fig:DynamicAmplitude}.  Due to the lack of any preferred direction in the changing permittivity \(\delta\alpha\) the emission is isotropic, but otherwise it is in many respects similar to that of a dielectric in oscillatory motion. However, note that the emission rate is not proportional $\nu^{2}$ in this case.
%
%
%
%
\subsection{Emission rate from a travelling refractive index perturbation}
\par
As a final, more involved example of macroscopic QED applied to time dependent media, we address the case of a medium through which a perturbation of the refractive index travels at a constant velocity $\mathbf{v}=v\hat{\mathbf{z}}$, producing pairs of polaritons. Recent work~\cite{AnalogHawking,FaccioAnalog,FaccioAnalog2} has established a connection between such a process and an analogue of Hawking radiation~\cite{AnalogHawking,HawkingRadiation}.  Here we do not emphasize the connection to general relativity, but rather look for a description of such an emission process that fully accounts for the effects of dispersion and dissipation.  To the authors' knowledge, previous treatments have not fully accounted for such effects.  We assume that the travelling perturbation to the permittivity can be represented in terms of the function $\delta\alpha(\omega,\mathbf{x},t)$ taking the form
\begin{equation}
	\delta\mathbf{\alpha}\left(\omega,\mathbf{x},t\right)=\beta\left(\omega\right)
\intop\frac{d^{3}\mathbf{k}}{\left(2\pi\right)^{3}}f\left(\mathbf{k}\right)e^{i\mathbf{k.}\left(\mathbf{x}-v\hat{\mathbf{z}}t\right)}
	\end{equation}
Unlike the case of a truly moving medium, $v$ is not restricted to non-relativistic velocities. In this case we find the following expression for the rate of change of the polariton amplitude,
\begin{equation}
	\dot{\boldsymbol{\zeta}}^{(b)}\left(\mathbf{x_{1}},\mathbf{x_{2}},\omega_{1},\omega_{2},t\right)=\intop\frac{d^{3}\mathbf{k}_{1}}{\left(2\pi\right)^{3}}\intop\frac{d^{3}\mathbf{k}_{2}}{\left(2\pi\right)^{3}}\boldsymbol{\mathcal{A}}\left(\mathbf{k}_{1},\mathbf{k}_{2},\omega_{1},\omega_{2}\right)f\left(\mathbf{k}_{1}+\mathbf{k}_{2}\right)e^{-i\left(\mathbf{k}_{1}+\mathbf{k}_{2}\right).\mathbf{v}t}e^{i\left(\omega_{1}+\omega_{2}\right)t}e^{i\mathbf{k}_{1}.\mathbf{x_{1}}}e^{i\mathbf{k}_{2}.\mathbf{x_{2}}}
\end{equation}
where the dyadic $\boldsymbol{\mathcal{A}}$ is given by
%
%
\begin{multline}
	\boldsymbol{\mathcal{A}}\left(\mathbf{k}_{1},\mathbf{k}_{2},\omega_{1},\omega_{2}\right)=-\frac{i\mu_{0}\alpha_{b}(\omega_{1})\omega_{1}^{2}}{4\sqrt{\omega_{1}\omega_{2}}}\mathbf{G}^{\dagger}(\mathbf{k}_{1},\omega_{1})\cdot\Bigg[\mu_{0}\omega_{2}^{2}\alpha_{b}(\omega_{2})\intop_{0}^{\infty}d\omega\frac{\beta(\omega)\alpha_{b}(\omega)}{\omega^{2}-(\omega_{2}-i\omega 0^{+})^{2}}\mathbf{G}^{\star}(\mathbf{k}_{2},\omega_{2})+\beta(\omega_{2})\mathbb{1}_{3}\bigg]\\
	+1\leftrightarrow2\label{AHawking}
\end{multline}
For simplicity, we choose the change in permittivity to have the same dispersion as the background material \(\beta(\omega)=\alpha_{b}(\omega)\), although presumably physically this need not be the case.  Equation (\ref{AHawking}) then reduces to
\begin{equation}
	\boldsymbol{\mathcal{A}}\left(\mathbf{k}_{1},\mathbf{k}_{2},\omega_{1},\omega_{2}\right)=-\frac{i}{2\pi}\sqrt{{\rm Im}[\chi\left(\omega_{1}\right)]{\rm Im}[\chi\left(\omega_{2}\right)]}\frac{\omega^{2}_{1}}{c^{2}}\mathbf{G}^{\dagger}\left(\mathbf{k}_{1},\omega_{1}\right)\cdot\left[\chi_{b}^{\star}\left(\omega_{2}\right)\frac{\omega^{2}_{2}}{c^{2}}\mathbf{G}^{*}\left(\mathbf{k}_{2},\omega_{2}\right)+\mathbb{1}_{3}\right]+1\leftrightarrow2.\label{eq:AHawking}
\end{equation}
Defining the net emission rate in terms of a spectral density that is now a function of two wave--vectors and frequency 
\[
	\Gamma=\intop_{0}^{\infty}d\omega\int\frac{d^{3}\mathbf{k}_{1}}{(2\pi)^{3}}\int\frac{d^{3}\mathbf{k}_{2}}{(2\pi)^{3}}\ \rho(\mathbf{k}_{1},\mathbf{k}_{2},\omega)
\]
we find \(\rho\) to be
\begin{equation}
	\rho(\mathbf{k}_{1},\mathbf{k}_{2},\omega)=2\pi\sum_{m,n}\left|\mathcal{A}_{mn}(\mathbf{k}_{1},\mathbf{k}_{2},\omega,\mathbf{v}\cdot(\mathbf{k}_{1}+\mathbf{k}_{2})-\omega)\right|^{2}\left|f(\mathbf{k}_{1}+\mathbf{k}_{2})\right|^{2}
\theta(\mathbf{v}\cdot(\mathbf{k}_{1}+\mathbf{k}_{2})-\omega)\label{eq:HawkingSpectralDensity}
\end{equation}
where we applied
\begin{equation}\label{eq:deltaAH}
	\underset{T\rightarrow\infty}{\lim}\frac{1}{T}\left|\intop_{-T/2}^{T/2}dt\ e^{i\left(\omega_{1}+\omega_{2}-\mathbf{v}\cdot(\mathbf{k}_{1}+\mathbf{k}_{2})\right)t}\right|^{2}=2\pi\delta\left[\omega_{1}+\omega_{2}-\mathbf{v}\cdot(\mathbf{k}_{1}+\mathbf{k}_{2})\right].
\end{equation}
%
%
The expression for the spectral density has a similar form to that given in the previous sections, except that it now depends on two wave vectors $\mathbf{k}_{1}$ and $\mathbf{k}_{2}$, due to the momentum exchanged between the polaritons and the moving perturbation; it also now has dimensions of $\mbox{L}^6$. In fact, the form of (\ref{eq:HawkingSpectralDensity}) can be motivated from fairly simple physical considerations: the spatial distribution of the moving perturbation \(\delta\alpha\) is determined by the function \(f(\mathbf{x})\), which has a corresponding Fourier spectrum \(f(\mathbf{K})\).  The two polaritons can exchange momentum with the moving perturbation so long as the conservation law \(\hbar\mathbf{K}=\hbar(\mathbf{k}_{1}+\mathbf{k}_{2})\) is satisfied.  Energy conservation must also be obeyed, with the pulse containing frequencies \(\mathbf{v}\cdot\mathbf{K}\) so that \(\hbar(\omega_{1}+\omega_{2})=\hbar\mathbf{v}\cdot\mathbf{K}\).  Combining these two conservation laws leads to \(\omega_{2}=\mathbf{v}\cdot(\mathbf{k}_{1}+\mathbf{k}_{2})-\omega_{1}\), with the rate of pair production being proportional to the amplitude of the relevant Fourier amplitude of the perturbation, \(|f(\mathbf{k}_{1}+\mathbf{k}_{2})|^{2}\).  The \(\theta\) function in (\ref{eq:HawkingSpectralDensity}) ensures that the energy of both of the polaritons is positive, while \(|\mathcal{A}_{nm}|^{2}\) scales the emission rate depending on how close the frequencies \(\omega_{1}\) and \(\omega_{2}\) are to the resonances of the material (where \(\alpha_{b}\) is large), and how close the dispersion relations \(|\mathbf{k}_{1,2}|^{2}=\epsilon^{\star}(\omega_{1,2})\omega_{1,2}^{2}/c^{2}\) are to being fulfilled (where the Green function \(\mathbf{G}\) is large).  From (\ref{eq:AHawking}) it is evident that the peak pair production will occur when the dispersion relation is fulfilled while the frequency of one member of the pair is at a resonance of the material.
\par
\begin{figure}[h]
		\includegraphics[height=7cm]{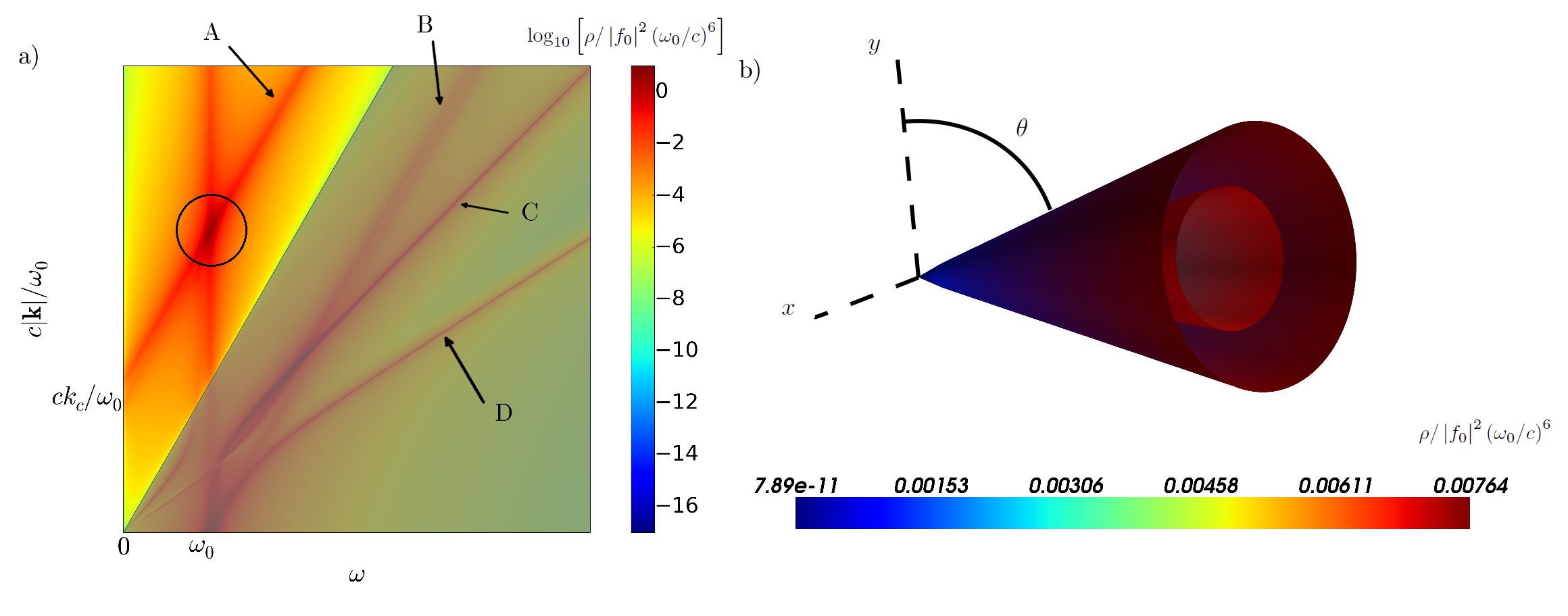}
		\caption{a): Logarithm of the spectral density for polariton emission $\log_{10}\left[\rho/|f_{0}|^{2}(\omega_{0}/c)^{6}\right]$ for an arbitrary inclination angle $\cos(\theta)=k_{z}/|\mathbf{k}|=1/\sqrt{3}$. The background dielectric function of the material \(\epsilon_{b}\) is given by the Lorentzian response (\ref{lorentzian}) with a resonance at $\omega_{0}$. The velocity is taken to be $v=0.5c$ and the Gaussian pulse width as $\sigma=0.1c/\omega_{0}$. The spectral density is significant at frequencies in the vicinity of the resonant frequency of the material (vertical line at  $\omega=\omega_{0}$), the shifted resonant frequencies $v(k_{z}+k'_{z})\pm\omega_{0}$ (A and B) and along the dispersion curve satisfying $k^{2}-\epsilon^{*}(\omega_{k})
\omega_{k}^{2}=0$ (see for example line D) and the shifted dispersion curve $k^{2}-\epsilon^{*}(\omega_{k,v}-vk_{z})(\omega_{k,v}-vk_{z})^{2}=0$ (see for example line C). The shaded area to the right of the line  $\omega=v(k_{z}+k'_{z})$ does not contribute to the total emission rate due to the theta function in the spectral density (\ref{dens}). The spectral density is greatest around intersections between dispersion curves and/or resonances of the material. The black circle points to an example of a region where two resonance curves intersect but similar intersections occur between resonances and dispersion curves or pairs of dispersion curves.   b) Angular dependence of the spectral density $\rho/|f_{0}|^{2}(\omega_{0}/c)^{6}$ for a fixed frequency $\omega_{1}=\omega_{0}$ where the frequency and wave vector magnitude are arbitrarily chosen to be $\omega=\omega_{0}$ and $|\mathbf{k}|=\omega_{0}\sqrt{3}$ respectively corresponding to the centre of the black circle in (a), for varying angle between $\mathbf{k'}$ and $\mathbf{k}$. The rate is seen to be negligible for all directions except along the surface of a number of cones traced out along rings of constant inclination angle $\theta$.}
		\label{fig:DynamicAmplitude_Full}
\end{figure}

In figure~\ref{fig:DynamicAmplitude_Full} we plot the dependence of \(\rho(\mathbf{k}_{1},\mathbf{k}_{2},\omega)\) as a function of both the relative angle between \(\mathbf{k}_{1}\) and \(\mathbf{k}_{2}\), and the magnitude of the modulus of both wave--vectors for a fixed angle between them.  We take the particular case where the moving perturbation to the permittivity takes the form of a Gaussian,
\[
	f(\mathbf{k})=f_{0}\exp\left(\frac{\sigma^{2}}{4}\mathbf{k}^{2}\right)
\]
with the area under the function \(f(\mathbf{x})\) equal to \(f_{0}\).  The spectral density for the excitation of polariton pairs (\ref{eq:HawkingSpectralDensity}) then becomes
\begin{equation}\label{dens}
	\rho(\mathbf{k}_{1},\mathbf{k}_{2},\omega)=2\pi |f_{0}|^{2}\sum_{m,n}\left|\mathcal{A}_{mn}(\mathbf{k}_{1},\mathbf{k}_{2},\omega,\mathbf{v}\cdot(\mathbf{k}_{1}+\mathbf{k}_{2})-\omega)\right|^{2}\exp\left(\frac{\sigma^{2}}{2}[\mathbf{k}_{1}+\mathbf{k}_{2}]^{2}\right)
\theta(\mathbf{v}\cdot(\mathbf{k}_{1}+\mathbf{k}_{2})-\omega)
\end{equation}
It is evident from (\ref{dens}) that there is a trade off in the spectral density between the requirement of energy conservation, and the Fourier spectrum of the moving perturbation.  The maximum in the Fourier spectrum \(f(\mathbf{k}_{1}+\mathbf{k}_{2})\) in this case occurs when \(\mathbf{k}_{1}=-\mathbf{k}_{2}\), which is where the energy of one of the polaritons \(\mathbf{v}\cdot(\mathbf{k}_{1}+\mathbf{k}_{2})-\omega\) is negative, and thus the spectral density is zero.  Meanwhile when \(\mathbf{k}_{1}=\mathbf{k}_{2}\), both polaritons can have positive energy, but the exponential factor \(\exp(\sigma^{2}[\mathbf{k}_{1}+\mathbf{k}_{2}]^{2}/2)\) reduces the spectral density exponentially with the magnitude of \(\mathbf{k}_{1,2}\).  Spatially sharp, quickly moving perturbations thus have relatively large emission rates for pairs of polaritons, and these pairs tend to be emitted away from the axis of propagation.  This is in broad agreement with e.g.~\cite{AnalogHawking} but here derived without the analogy to general relativity.  Spatially broad perturbations correspond to large values of \(\sigma\), and thus small values of \(\mathbf{k}_{1,2}\).  As the velocity of the perturbation drops to $0$, the \(\theta\) function in (\ref{dens}) picks out ever larger magnitudes of \(\mathbf{k}_{1,2}\), which are ever further from fulfilling the dispersion relation (i.e. where the Green function \(\mathbf{G}(\mathbf{k},\omega)\) in (\ref{eq:AHawking}) becomes ever smaller).  In this zero velocity limit the spectral density thus reduces to zero, as expected.
\par
%
%
%
%
\par
Figure~\ref{fig:DynamicAmplitude_Full}a confirms that the emission rate spectral density contains terms that contribute significantly for frequencies and wave-vectors where one of the pairs is close to fulfilling the dispersion relation and/or close to the resonant frequency of the material response.  This subfigure is plotted for an arbitrarily chosen angle between \(\mathbf{k}_{1}\) and \(\mathbf{k}_{2}\), with both wave--vectors having equal magnitude.  Figure~\ref{fig:DynamicAmplitude_Full}b represents the angular dependence of the spectral density of the emission, varying the angle between \(\mathbf{k}_{1}\) and \(\mathbf{k}_{2}\) for a fixed frequency and equal magnitude of the two wave--vectors.   The emission is concentrated in several cone-like regions. These cones correspond to either the shifted dispersion curves or the shifted resonances in figure \ref{fig:DynamicAmplitude_Full}a.  Increasing the absorption within the dielectric response has the effect of blurring the lines of \ref{fig:DynamicAmplitude_Full}a and of thickening the shells of the cone-like shapes of \ref{fig:DynamicAmplitude_Full}b.  This effectively means that it becomes possible to excite polaritons that lie further from the resonance of the material or the dispersion curve for electromagnetic waves.  The choices of plot in figure~\ref{fig:DynamicAmplitude_Full} do not encode the complete radiation pattern from the moving moving perturbation but serve to show under what conditions the emission of polariton pairs is increased, as well as the directional distribution of the emission.  It would be too lengthy to fully explain the different regimes of this effect here, and further results will be given in a future publication.
\par
It is interesting to compare this effect with those explored in sections~\ref{homed} and~\ref{sec:tdp}; in particular the time dependent permittivity of section~\ref{sec:tdp} which would typically be called a dynamic--Casimir type effect.  In cases of oscillatory motion or material time dependence, pairs of polaritons can be emitted with the total energy of the pair \(\hbar\nu\) coming from the motion.  We have found that so long as energy and momentum are conserved and the energy of both polaritons is positive, there is always some excitation in the material.  However, this is typically very small unless the wave--vectors and frequencies are close to either a material resonance, or to fulfilling the dispersion relation for electromagnetic waves.  The emission from a moving perturbation is of exactly the same form, and from the perspective of the perturbation theory comes from the same `dynamic Casimir' term (\ref{secondTerm}).  The only difference is that in this case the energy available for the creation of excitations instead comes from the frequencies \(\mathbf{v}\cdot\mathbf{k}\) within the moving perturbation, the \(\mathbf{k}\) dependence of which complicates the functional form of the emission spectral density.
\section{Discussion}
\par
The motivation behind this calculation was to understand the effects of the time dependence and motion of dielectric media on the excitations of polaritons (i.e. the normal modes of light and matter), when the effects of dissipation and dispersion are fully taken into account.  In particular we analysed the necessary modifications to the Hamiltonian of macroscopic QED that must be made to describe the motion and time dependence of a dielectric, and applied this to calculate the polariton excitation rate in the three illustrative cases summarized in figure~\ref{fig:System}.  Before calculating this emission rate we briefly discussed that the distinct effect of motion versus time dependence is to introduce non-locality into the permittivity.   In the case of uniform translational motion this non-locality is simply the well--known Doppler shift of the frequency dispersion, but that for more complicated motion becomes increasingly intricate. 
\par
Having found the separate modifications to the Hamiltonian of macroscopic QED necessary to describe moving or time dependent dielectric media, we applied time dependent perturbation theory to calculate the emission rate of polaritons in three particular cases.  In general we found that so long as the energies of both polaritons are positive, and energy and momentum are conserved then there is always some---albeit often very small---excitation rate.  For a uniform medium performing an oscillatory motion, or material properties that oscillate in time with angular frequency \(\nu\), energy conservation demands that the total energy of the pair equals \(\hbar\nu\).  For a material with a single resonance at \(\omega_{0}\) this means that there will be a peak in the emission (although not the only peak) when e.g. one of the pair has frequency \(\omega_{0}\) and wave--vector \(\mathbf{k}\), and the other has frequency \(\nu-\omega_{0}\) and wave--vector \(-\mathbf{k}\) where \(\mathbf{G}(-\mathbf{k},\nu-\omega_{0})\) is very large (i.e. close to fulfilling the dispersion relation).   Experimentally it might be of particular importance that in all the cases we examined, the regions where the polaritons both nearly fulfil the dispersion relation and are close to a material resonance contribute most to the emission.
\par
The final example we considered was the emission of polariton pairs from a uniformly moving perturbation to the permittivity of a medium.  We found that the description of this process is not fundamentally different from that of emission from a uniform time oscillating permittivity.  The main difference is that rather than take energy \(\hbar\nu\) from the oscillation of the material properties, the polaritons take energy \(\hbar\mathbf{v}\cdot\mathbf{k}\) from the motion of the perturbation.  The wave--vector dependence of the energy makes the emission increase with spatially sharper perturbations, and typically causes the polaritons to be emitted in cones.  However, the peak emission is still concentrated around the regions where the frequencies where one of the pairs is close to a material resonance, and the other is close to fulfilling the dispersion relation for electromagnetic waves, where both are at resonances, or where both are close to fulfilling the dispersion relation.  This finding that a emitted photon may be paired with an excitation of the material at a resonance may be useful for understanding some of the recent experiments on analogue Hawking emission in optical media.

\appendix

\section{Operator expansion coefficients\label{ApA}}
\par
From \cite{TomMacro}, the coefficients in the expansion of the field operators in terms of the polariton creation and annihilation operators are given by
\begin{equation}
\begin{split}
	&\mathbf{f}_{\mathbf{B}}\left(\mathbf{x},\mathbf{x_{1}},\omega_{1}\right)=-i\mu_{0}\sqrt{\frac{\hbar}{2\omega_{1}}}\omega_{1}\alpha_{b}\left(\mathbf{x_{1}},\omega_{1}\right)\boldsymbol{\nabla}\times\mathbf{G}\left(\mathbf{x},\mathbf{x_{1}},\omega_{1}\right)\\
	&\mathbf{f}_{\mathbf{X}}\left(\mathbf{x},\mathbf{x_{2}},\omega,\omega_{2}\right)=\sqrt{\frac{\hbar}{2\omega_{2}}}\left[\frac{\mu_{0}\omega_{2}^{2}\alpha_{b}\left(\mathbf{x},\omega\right)\alpha_{b}\left(\mathbf{x_{2}},\omega_{2}\right)}{\omega{}^{2}-\left(\omega_{2}+i\omega0^{+}\right)^{2}}\mathbf{G}\left(\mathbf{x},\mathbf{x_{2}},\omega_{2}\right)+\delta^{(3)}\left(\mathbf{x}-\mathbf{x_{2}}\right)\delta\left(\omega-\omega_{2}\right)\mathbb{1}_{3}\right]\\
	&\mathbf{f}_{\mathbf{E}}\left(\mathbf{x},\mathbf{x_{1}},\omega_{1}\right)=\mu_{0}\sqrt{\frac{\hbar}{2\omega_{1}}}\omega_{1}^{2}\alpha_{b}\left(\mathbf{x_{1}},\omega_{1}\right)\mathbf{G}\left(\mathbf{x},\mathbf{x_{1}},\omega_{1}\right)\\
	&\mathbf{f}_{\mathbf{A}}\left(\mathbf{x},\mathbf{x_{1}},\omega_{1}\right)=-\frac{i}{\omega}\left[\mathbf{f}_{\mathbf{E}}\left(\mathbf{x},\mathbf{x_{1}},\omega_{1}\right)\right]_{T}\\
	&\mathbf{f}_{\mathbf{\Pi_{A}}}\left(\mathbf{x},\mathbf{x_{1}},\omega_{1}\right)=-\epsilon_{0}\epsilon_{b}(\mathbf{x},\omega_{1})\mathbf{f}_{\mathbf{E}}\left(\mathbf{x},\mathbf{x_{1}},\omega_{1}\right)-\sqrt{\frac{\hbar}{2\omega_{1}}}\alpha_{b}(\mathbf{x},\omega_{1})\delta^{(3)}(\mathbf{x}-\mathbf{x_{1}})\\
	&\mathbf{f}_{\mathbf{\Pi_{X}}}\left(\mathbf{x},\mathbf{x_{1}},\omega,\omega_{1}\right)=-i\omega_{1}\mathbf{f}_{\mathbf{X}}\left(\mathbf{x},\mathbf{x_{1}},\omega,\omega_{1}\right)
\end{split}
\end{equation}
where $0^{+}$ is an infinitely small number serving to shift the coefficients off the real axis in the complex plane, the subscript $T$ denotes the transverse components, and the Green function obeys
\begin{equation}\label{eq:GreensFourier}
	\boldsymbol{\nabla}\boldsymbol{\times}\boldsymbol{\nabla}\boldsymbol{\times}\mathbf{G}(\mathbf{x},\mathbf{x}_{2},\omega_{1})-\frac{\omega_{1}^{2}}{c^{2}}\epsilon_{b}(\mathbf{x},\omega_{1})\mathbf{G}(\mathbf{x},\mathbf{x}_{2},\omega_{1})=\mathbb{1}_{3}\delta^{(3)}(\mathbf{x}-\mathbf{x}_{2})
\end{equation}
with the boundary condition that the waves are outgoing at infinity (i.e. the retarded Green function).  For the case of an infinite homogeneous medium with permittivity \(\epsilon_{b}\), the solution to (\ref{eq:GreensFourier}) is given by
\begin{equation}
	\mathbf{G}\left(\mathbf{x_{1}}-\mathbf{x_{2}},\omega_{1}\right)=\intop\frac{d^{3}\mathbf{k}}{\left(2\pi\right)^{3}}		\mathbf{G}\left(\mathbf{k},\omega_{1}\right)e^{i\mathbf{k}.\left(\mathbf{x_{1}}-\mathbf{x_{2}}\right)}=\intop\frac{d^{3}\mathbf{k}}{\left(2\pi\right)^{3}}\frac{\mathbf{k}\otimes\mathbf{k}-\epsilon_{b}\left(\omega_{1}\right)\left(\omega_{1}/c\right)^{2}\mathbb{1}_{3}}{\epsilon_{b}(\omega_{1})\left(\omega_{1}/c\right)^{2}\left(\epsilon_{b}\left(\omega_{1}\right)\omega_{1}^{2}/c^{2}-k^{2}\right)}e^{i\mathbf{k}.\left(\mathbf{x_{1}}-\mathbf{x_{2}}\right)}.
\end{equation} 
where a positive imaginary part for \({\rm Im}[\epsilon_{b}]\) picks out the retarded Green function.
%
%
\section{The probability amplitude for polariton emission in a moving, oscillating medium\label{ApB}}
\par
For the case of a infinite homogeneous oscillating medium of section \ref{homed}, the rate of change amplitude $\boldsymbol{\mathcal{A}}\left(\mathbf{k},\omega_{1},\omega_{2}\right)$ is 
\begin{equation}
	\boldsymbol{\mathcal{A}}\left(\mathbf{k},\omega_{1},\omega_{2}\right)=\frac{1}{2}\left[\boldsymbol{\mathcal{B}}\left(\mathbf{k},\omega_{1},\omega_{2}\right)+\boldsymbol{\mathcal{B}}\left(-\mathbf{k},\omega_{2},\omega_{1}\right)\right]
\end{equation}
where the quantity \(\boldsymbol{\mathcal{B}}\left(\mathbf{k},\omega_{1},\omega_{2}\right)\) is given by
\begin{multline*}
	\boldsymbol{\mathcal{B}}\left(\mathbf{k},\omega_{1},\omega_{2}\right)=k_{z}\sqrt{\epsilon_{I}\left(\omega_{1}\right)\epsilon_{I}\left(\omega_{2}\right)} a\left(\omega_{1},\omega_{2}\right)\mathbf{G}^{\dagger}\left(\mathbf{k},\omega_{1}\right)\cdot\mathbf{G}^{*}\left(\mathbf{-k},\omega_{2}\right)\\[7pt]
	+b\left(\omega_{1},\omega_{2}\right)\mathbf{G}^{*}\left(\mathbf{-k},\omega_{2}\right)+c\left(\omega_{1},\omega_{2}\right)\mathbf{G}^{\dagger}\left(\mathbf{k},\omega_{1}\right)+d\left(\omega_{1},\omega_{2}\right)\mathbb{1}_{3}
\\[5pt]
	+e\left(\omega_{1},\omega_{2}\right)\frac{1}{k_{z}} \left(\mathbf{G}^{\dagger}\left(\mathbf{k},\omega_{1}\right)\times\mathbf{k}\right)\cdot\left(\mathbf{\hat{z}}\times\mathbb{1}_{3}\right)\cdot\left(\chi\left(\omega_{2}\right)\frac{\omega_{2}^{2}}{c^{2}}\mathbf{G}^{*}\left(-\mathbf{k},\omega_{2}\right)+\mathbb{1}_{3}\right)
\end{multline*}
with the coefficients within this quantity given as
\begin{align}
a\left(\omega_{1},\omega_{2}\right)&=\frac{1}{\pi c^{4}}\omega_{1}^{3}\omega_{2}^{2}\left(\frac{\epsilon^{*}\left(\omega_{1}\right)}{\omega_{1}^{2}-\left(\omega_{2}-i\omega_{1}0^{+}\right)^{2}}+\frac{\epsilon^{*}\left(\omega_{2}\right)}{\omega_{2}^{2}-\left(\omega_{1}-i\omega_{2}0^{+}\right)^{2}}\right)\\
b\left(\omega_{1},\omega_{2}\right)&=\frac{\omega_{1}}{\pi c^{2}}\frac{\omega_{2}^{2}}{\omega_{1}^{2}-\left(\omega_{2}-i\omega_{1}0^{+}\right)^{2}}\\
c\left(\omega_{1},\omega_{2}\right)&=\frac{\omega_{1}}{\pi c^{2}}\frac{\omega_{1}^{2}}{\omega_{2}^{2}-\left(\omega_{1}-i\omega_{2}0^{+}\right)^{2}}\\
d\left(\omega_{1},\omega_{2}\right)&=\frac{1}{2\sqrt{\epsilon_{I}\left(\omega_{1}\right)\epsilon_{I}\left(\omega_{2}\right)}}\delta\left(\omega_{1}-\omega_{2}\right)\\
e\left(\omega_{1},\omega_{2}\right)&=\frac{i\omega_{1}}{\pi c^{2}}
\end{align}

\section{A proof of a delta function identity (\ref{eq:delta})\label{ApC}}
\par
We wish to evaluate the limit 
\begin{equation}\label{eq:deltaapp}
	\underset{T\rightarrow\infty}{\lim}\ \frac{1}{T}\left|\intop_{-T/2}^{T/2}dt\ v\left(t\right)e^{i\left(\omega_{1}+\omega_{2}\right)t}\right|^{2}
\end{equation}
where $\mathbf{v}(t)=z_{0}\nu\cos(\nu t)\mathbf{\hat{z}}$.  Performing the integral and taking the limit yields
\begin{equation}\label{eq:deltaapp2}
\begin{split}
\underset{T\rightarrow\infty}{\lim}&\ \frac{1}{T}\left|\intop_{-T/2}^{T/2}dt\ v\left(t\right)e^{i\left(\omega_{1}+\omega_{2}\right)t}\right|^{2}=\\&\frac{z_{0}^{2}\nu^{2}}{4}\left(2\pi\delta\left(\omega_{1}+\omega_{2}-\nu\right)+2\pi\delta\left(\omega_{1}+\omega_{2}+\nu\right)+8\ \underset{T\to\infty}{\lim}\frac{\cos\left[\nu T\right]-\cos\left[\left(\omega_{1}+\omega_{2}\right)T\right]}{\left[\left(\omega_{1}+\omega_{2}\right)^{2}-\nu^{2}\right]T}\right)
\end{split}
\end{equation}
where the first two terms are derived in the same way as (\ref{eq:timedep}). The third can be evaluated by considering that it will eventually be convoluted with a function of $\omega_{1}$ such as
\begin{equation}
	\intop_{-\infty}^{\infty}d\omega_{1}\ f(\omega_{1})\times\underset{T\to\infty}{\lim}\frac{\cos\left[\nu T\right]-\cos\left[\left(\omega_{1}+\omega_{2}\right)T\right]}{\left[\left(\omega_{1}+\omega_{2}\right)^{2}-\nu^{2}\right]T}.
\end{equation}
Over the the range $(-\infty,\infty)$, only a small region will contribute to the integral when $T$ is large, namely the region around $\nu=\pm(\omega_{1}+\omega_{2})$. This integral is therefore equivalent to the following expression
\begin{equation}
	\underset{T\to\infty}{\lim}\intop_{-\Delta}^{\Delta}d\eta\ f(\nu-\omega_{2}+\eta)\times\frac{\cos\left[\nu T\right]-\cos\left[\left(\nu+\eta\right)T\right]}{\left[\left(\nu+\eta\right)^{2}-\nu^{2}\right]T}.
\end{equation}
For large $T$, $\Delta$ will be small and as such we can treat $\eta$ as small and make the following approximation
\begin{equation}
	\underset{T\rightarrow\infty}{\lim}\ \frac{1}{T}\left|\intop_{-T/2}^{T/2}dt\ v\left(t\right)e^{i\left(\omega_{1}+\omega_{2}\right)t}\right|^{2}\approx\underset{T\to\infty}{\lim}\intop_{-\Delta}^{\Delta}d\eta\ f(\nu-\omega_{2}+\eta)\times\frac{\sin\left[\nu T\right]}{2\nu}.
\end{equation}
In the limit $T\to\infty$, the range $\Delta\to 0$ and as such the area under the integral above vanishes. We can thus ignore the third term of equation (\ref{eq:deltaapp2}).


\begin{thebibliography}{99}

\bibitem{CasimirPlates}H. B. G. Casimir,\emph{Koninkl. Ned. Adak. Wetenschap. Proc.} \textbf{51}, 793 (1948).
\bibitem{TomCasimir}T. G. Philbin,\emph{ New J. Phys.} \textbf{13} 063026, (2011).
\bibitem{KramersKronig}L. D. Landau, E. M. Lifshitz and L. P. Pitaevskii, \emph{Electrodynamics of Continuous Media}, Butterworth-Heinemann, (1984).
\bibitem{QuantumFriction}A. I. Volokitin and B. N. J. Persson, \emph{Phys. Rev. Lett.} \textbf{106}, 094502 (2011).
\bibitem{NoQuantumFriction}T. G. Philbin and U. Leonhardt, \emph{New J. Phys} \textbf{11}, 033035 (2009).
\bibitem{ShearingTheVacuum}J. B. Pendry, \emph{J. Phys. Cond.. Matter} \textbf{9}, 10301 (1997).
\bibitem{AnalogHawking}T. G. Phlibin, C. Kuklewicz, S. Robertson, S. Hill, F. K\"onig and U. Leonhardt, \emph{Science} \textbf{319}, 1367 (2008).
\bibitem{HawkingRadiation}S. M. Hawking, \emph{Nature} \textbf{248}, 30 (1974).
\bibitem{WaterAnalog}S, Weinfurtner, E. W. Tedford, M. C. J. Penrice, W. G. Unruh and G. A. Lawrence, \emph{Phys. Rev. Lett.} \textbf{106}, 021302 (2011).
\bibitem{FaccioAnalog}F. Belgiorno, S. L. Cacciatori, M. Clerici, V. Gorini, G. Ortenzi, L. Rizzi, E. Rubino, V. G. Sala and D. Faccio, \emph{Phys. Rev. Lett.} \textbf{105}, 203901 (2010).
\bibitem{FaccioAnalog2}D. Faccio, \emph{Clas. Quant. Grav.} \textbf{29}, 224009 (2012).
\bibitem{FactOrFiction}J. B. Pendry, \emph{New J. Phys.} \textbf{12} 068002 (2010).
\bibitem{UlfResponse}U. Leonhardt, \emph{New J. Phys.} \textbf{12} 068001 (2010).
\bibitem{HopfieldModel}J. J. Hopfield, \emph{Phys. Rev.} \textbf{122} 1555 (1958).
\bibitem{HuttnerBarnettModel}B. Huttner, S. M. Barnett,\emph{Phys. Rev.} A \textbf{46} 4306, (1992).
\bibitem{Suttorp2004} L. G. Suttorp and M. Wubs, \emph{Phys. Rev.} A \textbf{70}, 013816 (2004).
\bibitem{ScheelQED}S. Scheel and S. Y. Buhmann, \emph{Acta Physica Solvaca} \textbf{58}, 675 (2008).
\bibitem{AmooshModel}M. Amooshahi and F. Kheirandish,\emph{J Phys.} A \textbf{41}, 275402 (2008).
\bibitem{TomMacro}T. G. Philbin, \emph{New J. Phys.} \textbf{12}, 123008 (2010).
\bibitem{ConstitutiveEquations}S. A. R. Horsley, \emph{Phys. Rev.} A \textbf{84}, 063822 (2011).
\bibitem{MovingMedia}S. A. R. Horsley, \emph{Phys. Rev.} A \textbf{86}, 023830 (2012).
\bibitem{Dodonov2010} V. V. Dodonov, \emph{Physica Scripta} \textbf{82} 038105 (2010).
\bibitem{ObservationDynCas}C. M. Wilson, G. Johansson, A. Pourkabirian, J. R. Johansson, T. Duty, F. Nori and P. Delsing,\emph{Nature} \textbf{479}, 376 (2011).
\bibitem{LifshitzTheory}I. E. Dzyaloshinskii, E. M. Lifshitz and L. P. Pitaevskii, \emph{Sov. Phys. Uspekhi} \textbf{4} 153 (1961).
\bibitem{MathMeth} G. B. Arkfen, H. J. Weber and F. E. Harris, \emph{Mathematical Methods for Physicists}, Elsevier (2013).
\bibitem{QFT}S. Weinberg, \emph{The Quantum Theory of Fields}, Cambridge University Press (1995).
\bibitem{DynCas}S. A. Fulling, P. C. W. Davies,\emph{Proc. Roy. Soc.} A \textbf{348}, 393 (1975).
\bibitem{QED}V. B. Berestetskii, E. M. Lifshitz and L. P. Pitaevskii, \emph{Quantum Electrodynamics}, Butterworth-Heinemann (2008).

\end{thebibliography}
\end{document}